\documentclass[aps,prb,amsmath,superscriptaddress,preprint,bibnotes,longbibliography,floatfix]{revtex4-2}

\usepackage{mathptmx,newtxtext,newtxmath,xspace}
\usepackage{amsbsy,bm,bbold}
\usepackage{siunitx}
\usepackage{graphicx,color,xcolor}
\usepackage{blindtext}
\usepackage{dcolumn}
\usepackage{tabularx}
\usepackage{lineno}
\usepackage{lipsum}
\usepackage[colorlinks=true, 
            linkcolor=blue, 
            urlcolor=blue,
            citecolor=blue]{hyperref}

\newcommand{\iu}{{i\mkern1mu}}

\newcommand{\NiP}{Na$_{2}$BaNi(PO$_4$)$_{2}$}

\newcommand{\MgP}{Na$_{2}$BaMg(PO$_4$)$_{2}$}

\newcommand{\RNum}[1]{\uppercase\expandafter{\romannumeral #1\relax}}

\begin{document}

\title{Bose-Einstein condensation of a two-magnon bound state in a spin-one triangular lattice}

\author{
Jieming~Sheng$^{1,2,3,*}$,
Jia-Wei~Mei$^{1,4,5,*,\dagger}$,
Le~Wang$^{1,4,*}$,
Xiaoyu~Xu$^{6}$,
Wenrui~Jiang$^{1,4}$,
Lei~Xu$^{7}$,\\
Han~Ge$^{1}$,
Nan~Zhao$^{1}$,
Tiantian~Li$^{1}$,
Andrea~Candini$^{8}$,
Bin~Xi$^{9}$,
Jize~Zhao$^{10,11}$,
Ying~Fu$^{12}$,\\
Jiong~Yang$^{13}$,
Yuanzhu~Zhang$^{13}$,
Giorgio~Biasiol$^{14}$,
Shanmin~Wang$^{1}$,
Jinlong~Zhu$^{1,12}$,\\
Ping~Miao$^{15,16}$,
Xin~Tong$^{15,16}$,
Dapeng~Yu$^{4,1}$,
Richard~Mole$^{17}$,
Yi~Cui$^{6}$,
Long~Ma$^{18}$,\\
Zhitao~Zhang$^{18}$,
Zhongwen~Ouyang$^{19}$,
Wei~Tong$^{18}$,
Andrey~Podlesnyak$^{20}$,
Ling~Wang$^{7}$,\\
Feng~Ye$^{20}$,
Dehong~Yu$^{17,\dagger}$,
Weiqiang~Yu$^{6,21,\dagger}$,
Liusuo~Wu$^{1,12,3,5,\dagger}$,
Zhentao~Wang$^{22,23,7,\dagger}$\\
\small$^{1}$Department of Physics, Southern University of Science and Technology, Shenzhen 518055, China \\
\small$^{2}$School of Physical Sciences, Great Bay University and Great Bay Institute for Advanced Study, Dongguan 523000, China \\
\small$^{3}$Guangdong Provincial Key Laboratory of Advanced Thermoelectric Materials and Device Physics, Department of Physics,
Southern University of Science and Technology, Shenzhen, 518055, China\\
\small$^{4}$Shenzhen Institute for Quantum Science and Engineering, Shenzhen 518055, China \\
\small$^{5}$Shenzhen Key Laboratory of Advanced Quantum Functional Materials and Devices, Southern University of Science and Technology, Shenzhen 518055, China \\
\small$^{6}$ School of Physics and Beijing Key Laboratory of Optoelectronic Functional Materials and Micro-nano Devices, Renmin University of China, Beijing 100872, China \\
\small$^{7}$ School of Physics, Zhejiang University, Hangzhou 310058, China \\
\small$^{8}$ Institute of Organic Synthesis and Photoreactivity, National Research Council of Italy, ISOF-CNR, Via Gobetti 101, 40129 Bologna, Italy \\
\small$^{9}$ College of Physics Science and Technology, Yangzhou University, Yangzhou 225002, China \\
\small$^{10}$ School of Physical Science and Technology \& Key Laboratory for Magnetism and
Magnetic Materials of the MoE, Lanzhou University, Lanzhou 730000, China \\
\small$^{11}$ Lanzhou Center for Theoretical Physics and Key Laboratory of Theoretical
Physics of Gansu Province, Lanzhou University, Lanzhou 730000, China \\
\small$^{12}$ Quantum Science Center of Guangdong-Hong Kong-Macao Greater Bay Area (Guangdong), Shenzhen 518045, China \\
\small$^{13}$ Department of Chemistry, Southern University of Science and Technology, Shenzhen 518055, China \\
\small$^{14}$ IOM CNR, Laboratorio TASC, Area Science Park Basovizza, Trieste 34149, Italy \\
\small$^{15}$ Institute of High Energy Physics, Chinese Academy of Sciences (CAS), Beijing 100049, China \\
\small$^{16}$ Spallation Neutron Source Science Center (SNSSC), Dongguan 523803, China \\
\small$^{17}$ Australian Nuclear Science and Technology Organisation, Lucas Heights, New South Wales 2234, Australia \\
\small$^{18}$ Anhui Key Laboratory of Low-Energy Quantum Materials and Devices, High Magnetic Field Laboratory, Chinese Academy of Sciences, Hefei 230031, China \\
\small$^{19}$ Wuhan National High Magnetic Field Center and School of Physics, Huazhong University of Science and Technology, Wuhan 430074, China \\
\small$^{20}$ Neutron Scattering Division, Oak Ridge National Laboratory, Oak Ridge, Tennessee 37831, USA \\
\small$^{21}$ Key Laboratory of Quantum State Construction and Manipulation (Ministry of Education), Renmin University of China, Beijing 100872, China \\
\small$^{22}$ School of Physics and Astronomy, University of Minnesota, Minneapolis, Minnesota 55455, USA \\
\small$^{23}$ Center for Correlated Matter, Zhejiang University, Hangzhou 310058, China \\
\small$^\dagger$Corresponding authors: Jia-Wei~Mei (meijw@sustech.edu.cn); Dehong~Yu (Dehong.Yu@ansto.gov.au); Weiqiang~Yu (wqyu$\textunderscore$phy@ruc.edu.cn); Liusuo~Wu (wuls@sustech.edu.cn); Zhentao~Wang (ztwang@zju.edu.cn) \\
\small$^*$These authors contributed equally to this work.
}

\date{\today}

\maketitle

\textbf{In ordered magnets, the elementary excitations are spin waves (magnons), which obey Bose-Einstein statistics. 
Similarly to Cooper pairs in superconductors, magnons can be paired into bound states
under attractive interactions. 
The Zeeman coupling to a magnetic field is able to tune the particle density through a quantum critical point (QCP), beyond which a ``hidden order'' is predicted to exist.
Here we report direct observation of the Bose-Einstein condensation (BEC) of the two-magnon bound state in \NiP{}. 
Comprehensive thermodynamic measurements confirmed the two-dimensional BEC-QCP at the saturation field.
Inelastic neutron scattering experiments were performed to establish the microscopic model.
An exact solution revealed stable 2-magnon bound states that were further confirmed by electron spin resonance and nuclear magnetic resonance experiments, demonstrating that the QCP is due to the pair condensation and the phase below saturation field is likely the long-sought-after spin nematic phase.
}

Magnons are the elementary excitations of typical magnetically ordered systems, and their interactions often lead to a plethora of emergent phenomena. In his seminal work, Bethe pointed out the possibility of forming 2-magnon bound states in a spin chain~\cite{BetheH1931}. Later, this concept was generalized to higher dimensions by Wotis and Hanus~\cite{WortisM1963,HanusJ1963}. As an analogue of Cooper pairs in superconductors, it has been predicted theoretically that the BEC~\cite{BatyevEG1984,GiamarchiT2008,ZapfV2014_RMP} of 2-magnon bound states at $T=0$ corresponds to a quantum phase transition into a new state of matter~\cite{ChubukovAV1991_chiral,ShannonN2006,ZhitomirskyME2010}, often termed the spin nematic (SN) state~\cite{BlumeM1969,AndreevAF1984,PapanicolaouN1988}. The SN state is a type of ``hidden order'', which breaks the spin rotational symmetry while having zero dipolar magnetic moments.

To date, multi-magnon bound states have been observed in several systems, including 1D spin chains~\cite{TorranceJB1969_bound1,TorranceJB1969_bound2,HoogerbeetsR1984,ZvyaginSA2007}, triangular lattices (TLs)~\cite{KatsumataK2000,BaiX2021,LegrosA2021}, and cold atoms~\cite{FukuharaT2013}. However, affirmative experimental evidence of 2-magnon condensation and the associated SN phase remains elusive. While the application of a magnetic field appears to be one of the most promising paths to tune the system through this QCP, it is a challenging task due to the high saturation fields and/or rather complicated magnetic exchanges involved for the known candidate materials~\cite{SvistovLE2011,ButtgenN2014,OrlovaA2017,NawaK2014,NathR2008,BhartiyaVK2021,IshikawaH2015,JansonO2016,YoshidaM2017,KohamaY2019}.

Here, we report a two-dimensional BEC of a 2-magnon bound state in the recently discovered spin-1 TL insulating antiferromagnet Na$_2$BaNi(PO$_4$)$_2$~\cite{LiN2021,DingF2021}. 
The small magnetic exchanges of this compound result in a low saturation field ($\sim \qty{1.8}{\tesla}$), that allows an accurate extraction of the model parameters by inelastic neutron scattering (INS) experiment in the fully polarized (FP) phase. 
2-magnon bound states were found to be stable from the exact solution of the Lippmann-Schwinger equation. Confirmation via electron spin resonance (ESR) and nuclear magnetic resonance (NMR) measurements demonstrated that the QCP at saturation field originates from the BEC of a 2-magnon bound state, responsible for the SN phase below saturation.

\begin{figure*}[t!]
\includegraphics[width=0.99\textwidth]{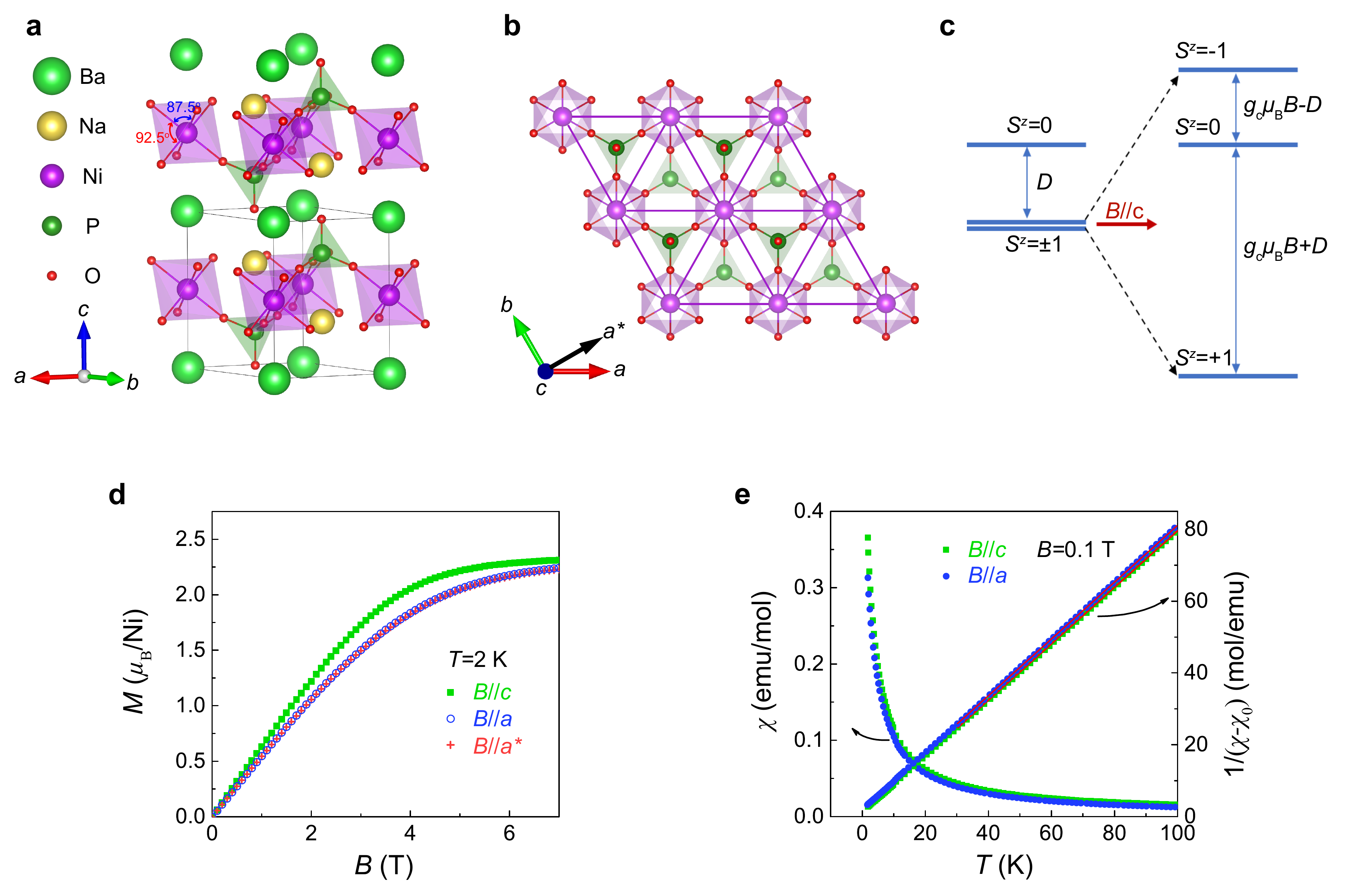}
\caption{ {\bf Crystal structure and magnetization of \NiP.} {\bf a-b} Crystal structure of \NiP{} from different viewing perspectives. 
The arrows \{$\bm{a}, \bm{b}, \bm{c}, \bm{a}^*$\} in the inset denote different high-symmetry directions.
{\bf c} Splitting of energy levels for the ground spin triplet state of Ni$^{2+}$ by crystal field and magnetic field in \NiP. {\bf d} Field-dependent magnetization measured at \qty{2}{K} with the magnetic field applied along $\bm{a}$, $\bm{a}^*$, and $\bm{c}$ directions. {\bf e} Temperature-dependent magnetic susceptibility $\chi$ (at $B=\qty{0.1}{T}$) and inverse susceptibility $1/{(\chi-\chi_0)}$, with a temperature-independent Van Vleck contribution $\chi_0$ removed.  The red solid lines are fits to the Curie-Weiss law, using data in the temperature range between \qty{30}{K} and \qty{300}{K}. }
\label{fig:structure}
\end{figure*}

Single crystals of \NiP{} were grown by the flux method~\cite{DingF2021,LiN2021}. Single-crystal X-ray diffraction data indicate that it crystallizes in a trigonal structure with space group $P\bar{3}m1$ (see Supplementary Information Sec.~\RNum{1}).
The crystal structure is shown in Figs.~\ref{fig:structure}{\bf a} and {\bf b}.
Magnetic Ni$^{2+}$ ions form a two-dimensional network of equilateral triangles in the $\bm{ab}$ plane, which are stacked in a simple A-A-A pattern along the $\bm{c}$-axis. 
In this refined crystal structure, the local point symmetry of Ni$^{2+}$ ions is $D_{3d} (\bar{3}m)$, resulting in a small trigonal distortion with the octahedra stretched along the $\bm{c}$-axis. As a $S=1$ spin system, the stretched trigonal distortion splits the ground spin triplet state into a lower $S^z=\pm1$ ground doublet state separated from the $S^z=0$ excited state by an energy $D$ that represents the easy-axis anisotropy. Applying a magnetic field $B$ along the easy axis will further split the ground doublet state, as shown in Fig.~\ref{fig:structure}{\bf c}.

Figure~\ref{fig:structure}{\bf d} shows the isothermal magnetization $M(B)$ measured at \qty{2}{K} with a magnetic field applied along the $\bm{c}$ axis ($B\parallel \bm{c}$) and in the $\bm{ab}$ plane ($B\parallel\bm{a}$, and $B\parallel\bm{a}^*$). For field in the $\bm{ab}$ plane, the magnetization is nearly isotropic with a saturation moment $\sim$ \qty{2.23}{\mu_B/Ni} at \qty{7}{T}. For field along the $\bm{c}$ axis, a slightly larger saturation moment $\sim$ \qty{2.31}{\mu_B/Ni} was observed at \qty{7}{T}. The smaller saturation field along the $\bm{c}$ aixs indicates that the anisotropy is of easy-axis type. The magnetic susceptibility $\chi(T)$ with $B \parallel \bm{a}$ and $B \parallel \bm{c}$ are presented in Fig.~\ref{fig:structure}{\bf e}, with no indication of a long-range order down to \qty{1.8}{K} for either field direction.
A Curie-Weiss fit of the inverse susceptibility yields $\theta_{\rm CW}^{ab} \simeq \qty{-2.23(3)}{K}$ and $\theta_{\rm CW}^{c} \simeq \qty{-1.22(4)}{K}$, with the Van Vleck contribution $\chi_0 =0.00326(1)$ for $B \parallel \bm{c}$ and $\chi_0 =0.00009(1)$ for $B \parallel \bm{a}$.  
The $g$-factors were extracted as $g_{ab}\simeq2.24$(1), $g_{c}\simeq2.25$(1) for $B \perp \bm{c}$ and $B \parallel \bm{c}$, respectively.

\begin{figure}[t!]
\includegraphics[width=\textwidth]{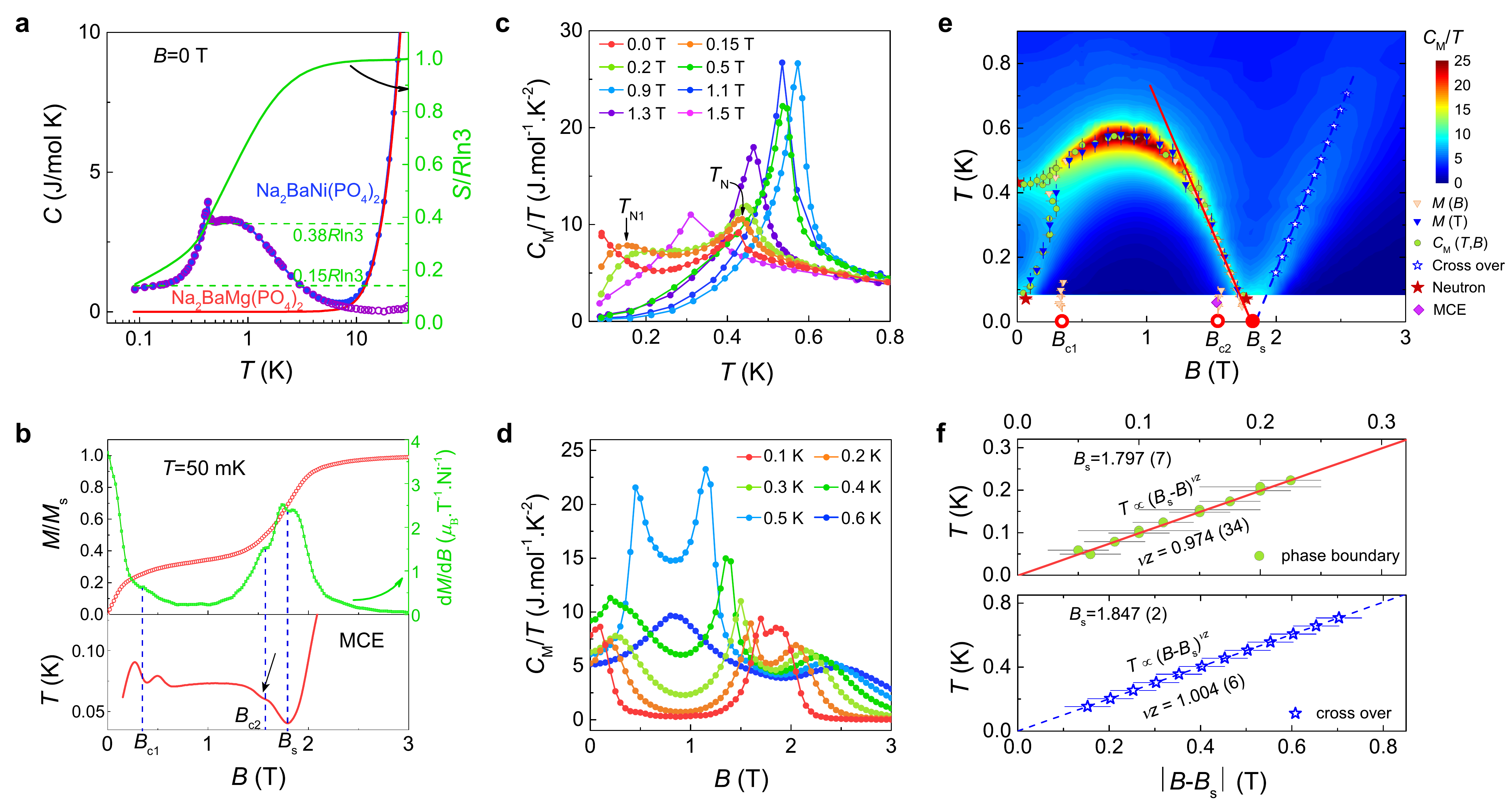}
\caption{{\bf Magnetic phase diagram and critical scaling behaviors of \NiP{}, with a field applied along the $\bm{c}$ axis.} 
{\bf a} Left axis: Temperature-dependent specific heat of~\NiP{} (blue filled circles) and~\MgP{} (red solid line). The magnetic specific heat $C_{\rm M}$ (violet empty circles) of \NiP{} was extracted by subtracting the phonon contribution of \MgP{}. Right axis: The integrated magnetic entropy (green solid line), with an extrapolated residual entropy of about $0.15 R\ln 3$ for the inaccessible temperature region.
{\bf b} Upper panel: Field dependence of normalized magnetization $M/M_{\rm s}$ (red empty circles) and the corresponding differential susceptibility $dM/dB$ (green dotted line) measured at \qty{50}{mK}. Weak peak-like anomalies were observed in $\mathrm{d}M/\mathrm{d}B$ at the phase boundaries, as indicated by the vertical dashed lines. Lower panel: MCE measured by monitoring $T$ while sweeping the magnetic field up with a fixed bath temperature $T=\qty{50}{mK}$. The black arrow indicates the weak anomaly observed around $B_{\rm c2}$ in the MCE curve. {\bf c} Temperature-dependent magnetic specific heat $C_{\rm M}/T$ measured at different magnetic fields.  {\bf d} Field-dependent specific heat $C_{\rm M}/T$ measured at different temperatures. 
{\bf e} Temperature-field phase diagram of \NiP{}, overplotted on the contour map of the magnetic specific heat $C_\text{M}/T$. The phase boundaries are extracted from the magnetization ($M$), specific heat ($C_\text{M}$), MCE and neutron scattering data. {\bf f} Linear-like power-law behaviors of the magnetic phase boundary $T\propto(B_{\rm s}-B)^{\nu z}$ and the crossover temperatures $T^{*}\propto(B-B_{\rm s})^{\nu z}$ in the vicinity of the critical field $B_{\rm s}$, indicating a two-dimensional universality class of $\nu=1/2$, and $z=2$. Error bars denote standard deviations in {\bf e} and {\bf f}.}
\label{fig:phd}
\end{figure}

To explore the magnetic ground state, lower temperature measurements were performed. 
The zero-field specific heat is shown in Fig.~\ref{fig:phd}{\bf a}. By subtracting the specific heat of the nonmagnetic isostructural compound Na$_2$BaMg(PO$_4$)$_2$ as an estimation of the lattice contribution, the magnetic specific heat $C_{\rm M}(T)$ of \NiP{} was obtained. A sharp peak can be observed at $T_\text{N} =\qty{0.43}{K}$, indicating the transition to a long-range ordered state.



The magnetization measured at \qty{50}{mK} with $B\parallel \bm{c}$ is presented in Fig.~\ref{fig:phd}{\bf b}. A $M/M_{\rm s}=1/3$ plateau exists in a wide field range, consistent with the AC susceptibility reported in Ref.~\cite{LiN2021}. This is also evidenced by the subtle features observed in the differential magnetic susceptibility $\mathrm{d}M/\mathrm{d}B$ (see Fig.~\ref{fig:phd}{\bf b}). The anomaly $B_{\rm c2}$  near \qty{1.56}{T}, also confirmed by the magnetocaloric effect (MCE) in Fig.~\ref{fig:phd}{\bf b}, indicates a new phase emerging in between the $1/3$-plateau and the FP state (see Supplementary Information Sec.~\RNum{7}).
The anomaly near \qty{1.9}{T} is the result of a gap opening inside the FP phase. By sweeping the magnetic field at different fixed temperatures and vice versa, we can map out the phase boundaries shown in Fig.~\ref{fig:phd}{\bf e} (see Supplementary Information Sec.~\RNum{2}).


The phase boundaries are further corroborated by the anomalies in the magnetic specific heat. 
With a small field along the $\bm{c}$-axis, another maximum in $C_{\rm M}/T$ becomes {evident} 
(see Fig.~\ref{fig:phd}{\bf c}, the maximum at $T_{\rm N1}$ for $B=\qty{0.15}{T}$). 
This lower-field transition can be better seen by plotting $C_{\rm M}/T$ versus $B$ at different fixed temperatures (Fig.~\ref{fig:phd}{\bf d}).
Finally, the transition to the FP phase is revealed as sharp peaks in Figs.~\ref{fig:phd}{\bf c}-{\bf d}, and the crossover behavior in the FP phase is seen as the broad maximum in $C_{\rm M}/T$ (Fig.~\ref{fig:phd}{\bf d}). 
The complete phase diagram is summarized in Fig.~\ref{fig:phd}{\bf e}, with the phase boundaries marked by various measurements performed. 

In the vicinity of a field-induced QCP~\cite{BreunigO2017},
both the crossover points ($T^*$) and the phase boundary ($T_{\rm c}$) scale linearly in a wide temperature range (Fig.~\ref{fig:phd}{\bf e}--{\bf f}):
\begin{equation}
T^{*}\propto (B-B_{\rm s})^{1.004(6)}, \quad T_{\rm c}\propto (B_{\rm s}-B)^{0.974(34)},
\end{equation}
corresponding to critical exponents $\nu z = 1$. 
The linear scaling behaviors from both sides point to a QCP near $B_{\rm s}\approx \qty{1.8}{T}$, consistent with a two-dimensional BEC of the magnetic quasiparticles ($d=2$, $\nu = \frac{1}{2}$, and $z=2$). 
For $d=2$, the scaling of the phase boundary ($T_{\rm c}$) is controlled by the Berezinskii-Kosterlitz-Thouless transition, correct up to a double logarithmic correction~\cite{FisherDS1988, ZapfV2014_RMP}. For temperatures comparable to the interlayer couplings (not accessible in this experiment), the scaling will eventually cross over to the $d=3$ behaviors: $T_{\rm c}\propto (B_{\rm s}-B)^{2/d}$ with $d=3$ and $T^* \propto (B-B_{\rm s})$. We note that since $d=2$ is the upper critical dimension, the mean-field scaling could also be used, which would lead to the same exponents. 

\begin{figure*}[t!]
\includegraphics[width=\textwidth]{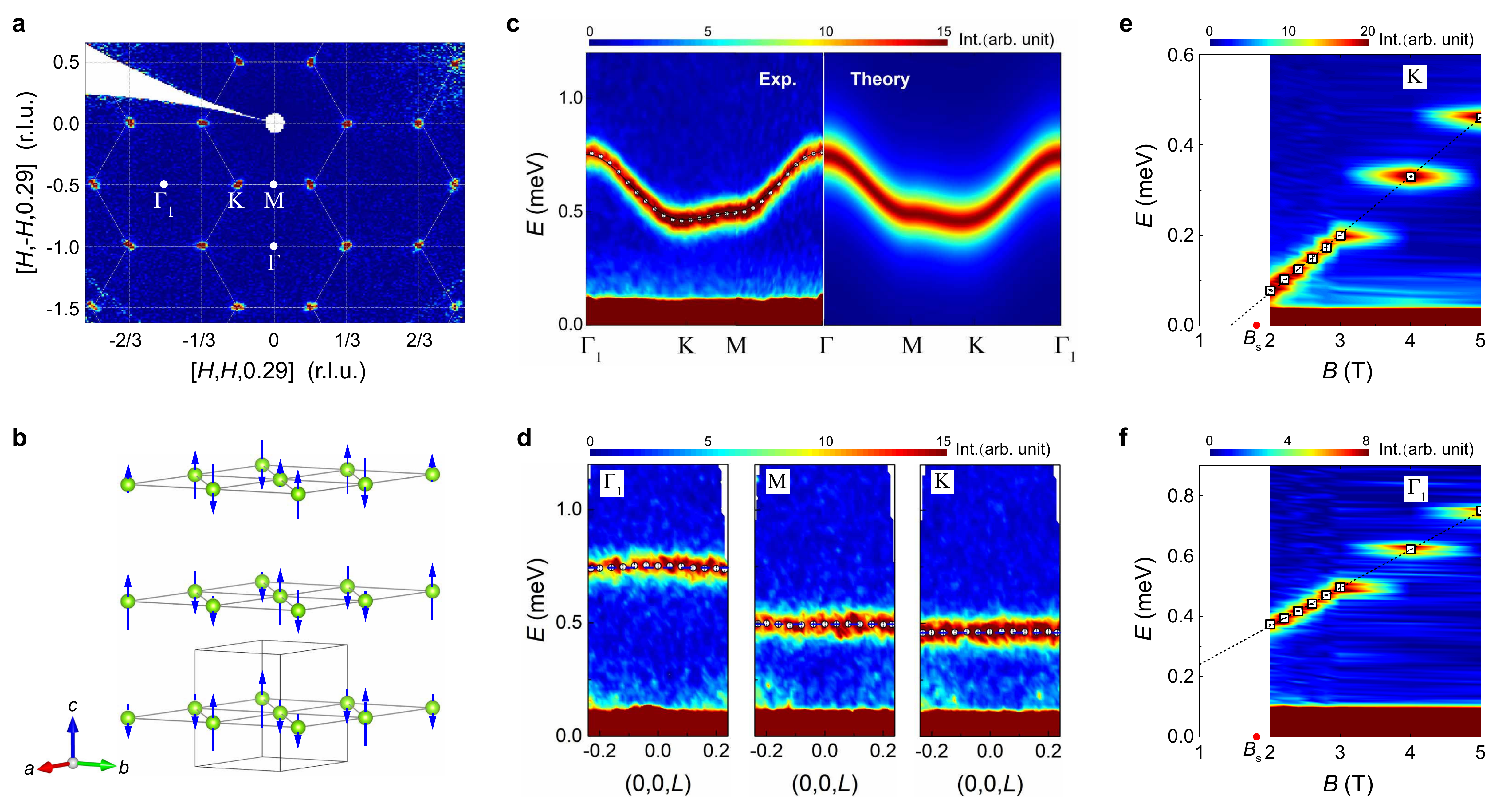}
\caption{{\bf Neutron scattering results of \NiP{}, with a field applied along the $\bm{c}$ axis.}
{\bf a} Neutron diffraction pattern measured in the $L=0.29$ plane at $T=\qty{80}{mK}$ (slightly above $T_{\rm N1}$) and $B=\qty{0}{T}$. {\bf b} The refined single-$Q$ magnetic structure for $T=\qty{80}{mK}$ and $B=\qty{0}{T}$. {\bf c} INS intensity measured at $T=\qty{60}{mK}$ and $B=\qty{5}{T}$ along high-symmetry directions. The intensity is integrated over the window of $L=[-0.22, 0.22]$. The black empty circles are the centers of the Gaussian fit to the data. The theoretical 1-magnon dispersion (black solid line on the left) and intensity (contour map on the right) are computed for the TL model~\eqref{eq:model_TL} with parameters given in Eq.~\eqref{eq:fit} (see Methods). {\bf d} INS intensity along the $L$-direction for different in-plane momenta at $T=\qty{60}{mK}$ and $B=\qty{5}{T}$. The black empty circles are centers of the Gaussian fit to the data and the blue dashed lines are guides to the eye. {\bf e}-{\bf f} Field dependence of the one-magnon excitations at the K point and the $\Gamma$ point, respectively. The black empty squares are the centers of the Gaussian fit to the one-magnon excitations at different magnetic fields (see Supplementary Information Sec.~\RNum{6}). The black dotted lines are linear fits of the one-magnon excitations at the K point and the $\Gamma$ point. Error bars denote standard deviations in {\bf c}-{\bf f}.}
\label{fig:neutron}
\end{figure*}

Single-crystal neutron scattering results are shown in Fig.~\ref{fig:neutron} (see Methods). 
At zero field and low temperature $T=\qty{80}{mK}$ (slightly above $T_{\rm N1}$), all the observed magnetic reflections can be classified by four inequivalent propagation vectors $\bm{k}_1=-\bm{k}_4=(1/3,\,1/3,\, \delta)$ and $\bm{k}_2=-\bm{k}_3=(1/3,1/3,-\delta )$ where $\delta \approx 0.293(1)$ (see Figs.~\ref{fig:neutron}{\bf a} and Fig.~S11 in Supplementary Information), indicating that the dominant in-plane interaction is the nearest-neighbor exchange. While inter-plane interactions are weak compared to the in-plane ones, the incommensurate component $\delta$ indicates that they are frustrated.

The zero-field magnetic structure at \qty{80}{mK} is determined by group theory analysis~\cite{Rodriguez-CarvajalJ1993} 
(see Supplementary Information Sec.~\RNum{4}).
For the space group $P\bar{3}m1$ with the Ni site at $(0,\,0,\,0.5)$, three irreducible representations (IRs) including \{$\Gamma_1$, $\Gamma_2$, $\Gamma_3$\} are the candidate single-$Q$ states for $\bm{k}= \bm{k}_1$, among which $\Gamma_1$ corresponds to the optimized state from refinement. 
As illustrated in Fig.~\ref{fig:neutron}{\bf b}, the spins are modulated along the $\bm{c}$ axis, forming an ``up-up-down'' spin structure in the $\bm{ab}$ plane. 
Although in this work we do not have lower-temperature neutron diffraction data, according to Fig.~\ref{fig:phd}{\bf e} and the theoretical analysis presented later, it is plausible that the magnetic structure goes through a sequence of transitions upon lowering temperature: at $T_\text{N1}$ the wave vector locks to $(1/3,\, 1/3,\, 1/3)$ that leads to the $\frac{1}{3}$-plateau phase, and at even lower temperature a spontaneous $U(1)$-symmetry breaking leads to a SN-supersolid phase.

From the above analysis, we have concluded that \NiP{} can be described by a quasi-2D spin-1 Hamiltonian dominated by the in-plane nearest-neighbor interactions. The space group $P\bar{3}m1$ puts restrictions on the in-plane nearest-neighbor interactions, 
where antisymmetric exchanges are forbidden. While other anisotropies are allowed across the nearest-neighbor bonds, we can safely ignore them because
the magnetization curves are practically identical for $B \parallel \bm{a}$ and $B \parallel 
\bm{a}^*$, and because of the $U(1)$ symmetry revealed by the scaling analysis with dynamical exponent $z=2$. 
The minimal model can be written as
\begin{equation}
\mathcal{H}= \mathcal{H}_\text{TL} + \mathcal{H}_c,
\label{eq:model_full}
\end{equation}
where the dominant in-plane Hamiltonian is the TL XXZ model involving only nearest-neighbor bonds $\langle ij \rangle_{ab}$:
\begin{equation}
\mathcal{H}_\text{TL}= J \sum_{\langle ij \rangle_{ab}} \left( S_i^x S_j^x + S_i^y S_j^y + \Delta S_i^z S_j^z \right)  -D \sum_i \left( S_i^z \right)^2.
\label{eq:model_TL}
\end{equation}
Two types of anisotropies are allowed by $U(1)$ symmetry: the exchange anisotropy $\Delta S_i^z S_j^z$ and the single-ion anisotropy $D \left( S_i^z \right)^2$.
The subdominant out-of-plane Hamiltonian $H_c$ is relatively complicated, and possibly influenced by the dipole-dipole interaction. To capture the incommensurate $L$-component of the ordering wavevector, multiple competing interactions should be present. Combined with the strong easy-axis single-ion anisotropy ($D/J \gg 1$), $\mathcal{H}_c$ should exhibit the qualitative physics of the ANNNI model~\cite{FisherME1980}, with the consequent incommensurate-commensurate transition at $T_{\rm N1}$ (see Fig.~\ref{fig:phd}{\bf e} and Supplementary Information Fig.~S11).

An accurate extraction of the model parameters can be achieved by performing INS experiment in the FP state. Figure~\ref{fig:neutron}{\bf c} shows the 
INS spectra along a high-symmetry momentum trajectory measured with $B=\qty{5}{T}$ 
applied along the $\bm{c}$-axis at $T=\qty{60}{mK}$. In this FP state, the 1-magnon excitation shows a clear dispersion in the $\bm{ab}$ plane with bandwidth $\sim \qty{0.3}{meV}$. 
In contrast, the magnetic excitation along the $L$-direction is almost flat (Fig.~\ref{fig:neutron}{\bf d}), confirming the quasi-2D nature of the compound: $\mathcal{H}_c \ll \mathcal{H}_\text{TL}$. 

The $U(1)$ symmetry of the TL model \eqref{eq:model_TL} is preserved when the magnetic field is applied along the $\bm{c}$-axis, allowing for an exact solution of the 1-magnon dispersion (see Methods):
\begin{equation}
E_1(\bm{k}) = 2J \left( \cos k_x + 2 \cos \frac{k_x}{2} \cos \frac{\sqrt{3}k_y}{2} \right) - 6 \Delta J + D + g_c \mu_B B.
\label{eq:1magnon}
\end{equation}

By fitting the INS data to Eq.~\eqref{eq:1magnon}, we obtain
\begin{equation}
J\approx \qty{0.032(1)}{meV},\quad g_c \approx 2.24(1),\quad -6\Delta J+D \approx \qty{-0.090(1)}{meV}.
\label{eq:fit}
\end{equation}
Inclusion of extra in-plane second-nearest-neighbor interaction $J_2$ yields $J_2\approx \qty{-0.0001(4)}{meV}$, again confirming that it is negligible.

While the parameters \{$\Delta$, $D$\} are still unknown, we can already determine the critical field corresponding to a ``would-be'' 1-magnon BEC at K-point (see also Fig.~\ref{fig:neutron}{\bf e}):
\begin{equation}
B_\text{1-magnon} = \frac{3J+6\Delta J - D}{g_c \mu_B} \approx \qty{1.43}{T}.
\end{equation}
Clearly, $B_\text{1-magnon}$ is well below the experimentally observed critical point $B_{\rm s}\approx \qty{1.8}{T}$. In other words, another transition overtakes the 1-magnon condensation as we lower the magnetic field.

Although we have only observed the 1-magnon state in the INS data (Fig.~\ref{fig:neutron}{\bf c}-{\bf f}), it is crucial to note that the multi-magnon states are invisible due to the $U(1)$ symmetry of the TL model \eqref{eq:model_TL}.
This becomes clear by considering the matrix elements of the $T=0$ dynamic spin structure factor:
\begin{equation}
\langle \nu^{(n)} | S_{-{\bm{k}}}^\alpha | \text{FP} \rangle \equiv 0,\quad  \forall n>1
\label{eq:matrix_element}
\end{equation}
since the good quantum number $\sum_i S_i^z$ in the $n$-magnon states $|\nu^{(n)} \rangle$ differs from that of the FP state by $n$-spin flips, which cannot be connected by a single operator $S_{-{\bm{k}}}^\alpha$ ($\alpha = x,y,z$).
In fact, the higher in-plane saturation field (Fig.~\ref{fig:structure}{\bf d}) indicates an easy-axis single-ion anisotropy ($D>0$), 
allowing the 2-magnon state to further reduce its energy by forming a single-ion bound state~\cite{SilberglittR1970}.

\begin{figure*}[t!]
\includegraphics[width=\textwidth]{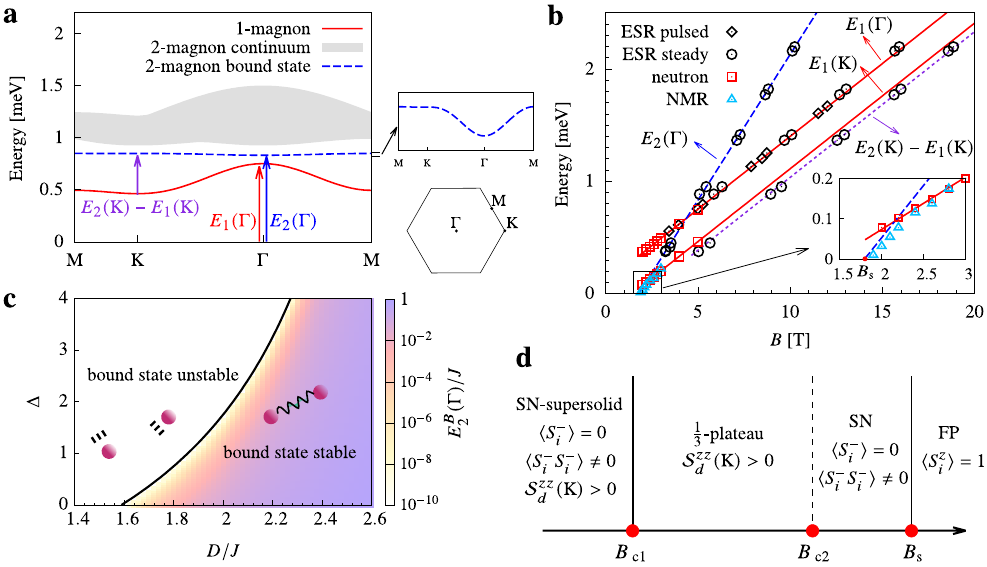}
\caption{{\bf Solution of the spin-1 model along with neutron, ESR and NMR data.}
{\bf a} Energies of the 1-magnon excitation, 2-magnon continuum and 2-magnon bound state at $B=\qty{5}{T}$ applied along the $\bm{c}$ axis, for the TL model~\eqref{eq:model_TL} with $J=\qty{0.032}{meV}$, $\Delta=1.13$, $D/J=3.97$, and $g_{c}=2.24$.
The arrows denote the excitations observed in our ESR experiments.
The inset shows the zoomed-in view of the 2-magnon bound state dispersion and the schematic plot of the first Brillouin zone. {\bf b} Energies of the magnon excitations as a function of the magnetic field, where the parameters are the same as in {\bf a} (the lines are solutions of the model, instead of being guides to the eye). The neutron scattering data measured at $T=\qty{60}{mK}$ in Figs.~\ref{fig:neutron}{\bf e}-{\bf f}, the ESR data measured at $T=\qty{2}{K}$ and the NMR data measured down to \qty{30}{mK} (see Supplementary Information Sec.~\RNum{8}-\RNum{9}) are also included for comparison. 
{\bf c} Stable region of the 2-magnon bound state in the FP state, indicated by positive binding energy $E_2^B(\Gamma)>0$.
{\bf d} Schematic $T=0$ phase diagram obtained by projecting $\mathcal{H}_\text{TL}$ to the hard-core Bose-Hubbard model \eqref{eq:boson}, where the solid (dashed) vertical lines indicate second (first) order transitions, and $\mathcal{S}_d^{zz}(\text{K})$ is the $z$-component of the dipolar spin structure factor at K point. According to Ref.~\cite{WesselS2005}, we can estimate $B_{\rm c1}\approx \qty{0.08}{T}$, $B_{\rm c2}\approx \qty{1.78}{T}$, and $B_{\rm s}\approx \qty{1.95}{T}$.}
\label{fig:2magnon}
\end{figure*}

To explore the possible formation of the 2-magnon bound state, we performed exact calculations by solving the Lippmann-Schwinger equation (see Methods). Figure~\ref{fig:2magnon}{\bf a} shows the 1-magnon and 2-magnon energies of the TL model \eqref{eq:model_TL} with $\Delta=1.13$ [$D/J\approx 3.97$ according to Eq.~\eqref{eq:fit}]. Besides the 1-magnon state and the 2-magnon continuum, the 2-magnon bound state is indeed found to be stable theoretically. The narrow band of the two-magnon bound state has a minimum at the $\Gamma$ point (see inset of Fig.~\ref{fig:2magnon}{\bf a}), suggesting it results from binding two magnons with opposite momenta. 
The binding energy is defined as $E_2^B(\bm{K})\equiv \left[E_1 \left(\frac{\bm{K}}{2} + \bm{k} \right) + E_1 \left(\frac{\bm{K}}{2} - \bm{k} \right) \right]_{\text{min }\bm{k}} - E_2(\bm{K})$ where $\bm{K}$ is the center-of-mass  momentum. Since $E_2^B(\bm{K}=\Gamma)$ increases against $D/J$ (Fig.~\ref{fig:2magnon}{\bf c}), 
it indicates that the easy-axis single-ion anisotropy provides the glue of the 2-magnon bound state.

As we reduce the magnetic field, the energies of the 1-magnon and 2-magnon states decrease with different slopes (Fig.~\ref{fig:2magnon}{\bf b}). With $\Delta=1.13$ [$D/J\approx 3.97$ according to Eq.~\eqref{eq:fit}], the 2-magnon bound state is found to condense at \qty{1.8}{T}, before the 1-magnon BEC could take place. This matches perfectly the experimentally observed QCP of $B_{\rm s}\approx \qty{1.8}{T}$, suggesting that it corresponds to a transition from the FP state to a SN state where $\langle S_i^{x} \rangle = \langle S_i^{y} \rangle = 0$ while $\langle S_i^{-} S_i^{-}\rangle \neq 0$. 
This SN state is ferro-nematic since the condensation occurs at the $\Gamma$ point.
We note that, for this parameter set the 3- and 4-magnon bound states are unstable in the TL model \eqref{eq:model_TL}, so there is no phase separation (potential 1st-order transition) to be concerned (see Supplemental Information Sec.~\RNum{9}).

To reveal the 2-magnon bound state, we performed steady-field ESR measurement with a magnetic field tilted $6^\circ$ away from the $\bm{c}$-axis to break the $U(1)$ symmetry.
Under such a configuration,
the intermediate state $|\nu^{(n)}\rangle$ in Eq.~\eqref{eq:matrix_element} necessarily includes a fraction of the 2-magnon contributions.
Two excitations associated with the 2-magnon bound states were revealed (see Fig.~\ref{fig:2magnon}{\bf b} and Supplementary Information Sec.~\RNum{8}). The first excitation agrees perfectly with the theoretically predicted values $E_2(\Gamma)$. The second excitation lies along the line with energy $E_2(\text{K})-E_1(\text{K})$, corresponding to a transition from the 1-magnon band to the 2-magnon bound state (finite-temperature effect). 

To obtain the 2-magnon bound state energy at even lower fields, we performed the $^{23}$Na nuclear magnetic resonance (NMR) measurements in the FP phase at temperatures down to \qty{30}{mK}. From the NMR spin-lattice relaxation rate $1/T_1$ (see Supplementary Information Sec.~\RNum{9}), we extracted the field dependence of the magnetic excitation gap.
Figure~\ref{fig:2magnon}{\bf b} shows the gap evolution for energies below \qty{0.2}{meV}. The slope change near \qty{2.2}{T} confirms the theoretical prediction that the lowest-energy excitation changes from the 1-magnon state to the 2-magnon bound state. More importantly, the gap closing indeed happens near $B_\text{s}$.
The direct experimental observation of 2-magnon bound states and the perfect agreement with theoretical predictions established firm evidence of the 2-magnon BEC at the critical field. As a result, the phase right below $B_{\rm s}$ should be SN~\cite{ChubukovAV1991_chiral,ShannonN2006,ZhitomirskyME2010}.

Since the combination of neutron, ESR and NMR data completely determined the parameters of the model \eqref{eq:model_TL}, it can be used reliably to analyze the various phases below $B_{\rm s}$. 
By expanding in the small parameter $J/D\sim 1/4$,
we can project the TL model~\eqref{eq:model_TL} to the $|\pm 1\rangle$ doublet space:
\begin{equation}
\mathcal{H}_\text{eff} = -\frac{J^2}{D} \sum_{\langle ij\rangle} \left[ s_i^x s_j^x + s_i^y s_j^y  -2 \left( 1 + \frac{2 \Delta D}{J} \right)  s_i^z s_j^z \right] - 2g_c \mu_B B \sum_i s_i^z,
\label{eq:spin-half}
\end{equation}
where $s_i^\alpha$ are the effective spin-$\frac{1}{2}$ operators. In terms of hard-core bosons~\cite{MatsubaraT1956}: $s_i^+=b_i$, $s_i^-=b_i^\dagger$, $s_i^z=1/2-b_i^\dagger b_i$, we obtain
\begin{equation}
\mathcal{H}_\text{eff} = -t\sum_{\langle ij\rangle} \left( b_i^\dagger b_j + h.c. \right) + V \sum_{\langle ij \rangle} n_i n_j - \mu \sum_i n_i,
\label{eq:boson}
\end{equation}
where $t=\frac{J^{2}}{2D}$, $V=\frac{2J^{2}}{D}+4\Delta J$, and $\mu = \frac{6J^{2}}{D}+12\Delta J-2g_{c}\mu_{B}B$. Substituting with $J=\qty{0.032}{meV}$, $\Delta=1.13$, and $D/J=3.97$, we obtain $t/V\approx 0.025$, $\mu/V = 3$ for $B=\qty{0}{T}$, and $\mu/V\approx 0.096$ for $B=\qty{1.8}{T}$.

In fact, the $T=0$ phase diagram of the low-energy effective model \eqref{eq:boson} was already known~\cite{WesselS2005,HeidarianD2005,MelkoRG2005,BoninsegniM2005}. 
At small $t/V$,  it consists of a supersolid phase for $\mu/V\approx 3$, a solid $\rho=1/3$ phase ($\rho$ is the boson density) that occupies a wide range of $\mu/V$ below 3, a superfluid phase near $\mu/V\approx 0$, and an empty phase for $\mu<-6t$. The solid $\rho =1/3$ phase is nothing but the $\frac{1}{3}$-plateau in both the effective spin-$\frac{1}{2}$ and the original spin-$1$ models. Since $s_i^-=P_0 \left( S_i^-S_i^-/2\right) P_0$ where $P_0$ is the projection operator to the doublet space, the superfluid phase of the effective spin-$\frac{1}{2}$ model is indeed the SN phase of the spin-$1$ model ($\langle S_i^-S_i^- \rangle \neq 0$). The supersolid phase includes not only long-range diagonal order in $s_i^z$, but also long-range off-diagonal order in terms of $s_i^-$. In other words, the superfluid order inside this supersolid phase is also SN in the original spin-$1$ model. This mapping establishes Fig.~\ref{fig:2magnon}{\bf d} that summarizes the $T=0$ phase diagram, that agrees well with the experiment Fig.~\ref{fig:phd}{\bf e}.

\begin{figure*}[t!]
\includegraphics[width=\textwidth]{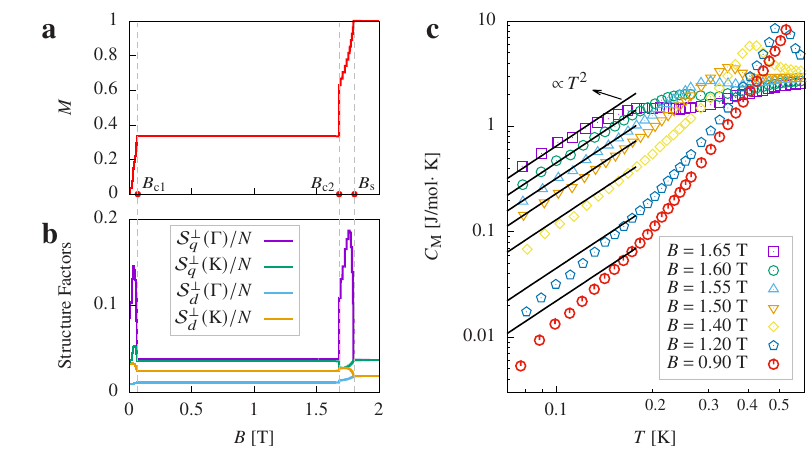}
\caption{{\bf Further evidence of the spin nematic phases.}
{\bf a} Magnetization, {\bf b} In-plane dipolar ($\mathcal{S}^\perp_d$) 
 and quadrupolar ($\mathcal{S}^\perp_q$) structure factors at $T=0$ as a function of field applied along the $\bm{c}$ axis, for the TL model \eqref{eq:model_TL} with $J=\qty{0.032}{meV}$, $\Delta=1.13$, $D/J=3.97$, and $g_{c}=2.24$, computed by DMRG on $N=9\times 6$ TL torus with bond dimension 3500.  {\bf c} Magnetic specific heat of \NiP{} measured at low temperature, where the straight lines are proportional to $T^2$ as guides to the eye.}
\label{fig:dmrg}
\end{figure*}

The zero-field phases of a similar spin-1 model were studied using the cluster mean-field (CMF) method ~\cite{Moreno-CardonerM2014}, which did not reach convergent results in the region where $D/J=3.97$. To improve on top of the perturbative and the CMF results, we performed density matrix renormalization group (DMRG) calculation~\cite{WhiteSR1992} on the TL model~\eqref{eq:model_TL}. The $T=0$ results with periodic boundary conditions along both directions are presented in Fig.~\ref{fig:dmrg}{\bf a}-{\bf b} (see Supplementary Information Sec.~\RNum{11} for additional results with cylindrical boundary condition), where we computed the magnetization, the in-plane dipolar and quadrupolar structure factors:
\begin{subequations}
\begin{align}
\mathcal{S}^\perp_{d}(\bm{k}) &\equiv \frac{1}{N} \sum_{i,j} e^{-\iu \bm{k} \cdot \left( \bm{r}_i - \bm{r}_j \right)} \langle S_i^x S_j^x + S_i^y S_j^y \rangle,\\
\mathcal{S}^\perp_{q}(\bm{k}) &\equiv \frac{1}{N} \sum_{i,j} e^{-\iu \bm{k} \cdot \left( \bm{r}_i - \bm{r}_j \right)} \langle Q_i^{x^2-y^2} Q_j^{x^2-y^2} + Q_i^{xy} Q_j^{xy} \rangle,
\end{align}
\end{subequations}
where $Q_i^{x^2-y^2} \equiv S_i^x S_i^x - S_i^y S_i^y$, $Q_i^{xy}\equiv S_i^xS_i^y + S_i^y S_i^x$, and $N$ is the total number of lattice sites.
The magnetization profile clearly indicates three critical points \{$B_{\rm c1}$, $B_{\rm c2}$, $B_{\rm s}$\}, where $B_{\rm s}\approx \qty{1.8}{T}$ which agrees well with the experiment. Note that $B_{\rm c1}$ ($B_{\rm c2}$) is a second (first) order transition, that also agrees with the perturbative calculation. For $B<B_{\rm c1}$ and $B_{\rm c2}< B < B_{\rm s}$, we see that the ferroquadrupolar order is indeed dominant in these putative SN phases. The in-plane dipolar structure factors $\mathcal{S}^\perp_{d}(\Gamma)$ and $\mathcal{S}^\perp_{d}(\text{K})$ are negligible in these SN phases, and are expected to be zero in the thermodynamic limit according to the perturbative analysis.
In contrast, the out-of-plane dipolar structure factor $\mathcal{S}^{zz}_{d}(\text{K})$ is nonzero for $B<B_\text{c2}$ (see also Fig.~S26 in Supplementary Information). For the SN-supersolid phase with $B<B_\text{c1}$, it is characterized by two nonzero structure factors: $\mathcal{S}^\perp_{q}(\Gamma)$ and $\mathcal{S}^{zz}_{d}(\text{K})$.


Since the SN order spontaneously breaks the $U(1)$ symmetry, a Goldstone mode is expected to appear~\cite{ChubukovAV1990,TsunetsuguH2006,LauchliA2006,MichaudF2011,SmeraldA2015}. Figure~\ref{fig:dmrg}{\bf c} shows the magnetic specific heat at low temperature: For $B_{\rm c2} < B < B_{\rm s}$, the $T^2$-behavior is indeed consistent with a linearly dispersive Goldstone mode in 2D; in contrast, the specific heat in the 1/3-plateau decreases much faster due to the gapped magnon spectra. While this $T^2$ specific heat behavior is consistent with the existing Goldstone mode, it is still an indirect one of the SN phase, and further measurements including ESR~\cite{FuruyaSC2018}, NMR~\cite{OrlovaA2017}, Raman~\cite{MichaudF2011} and neutron scattering~\cite{TsunetsuguH2006,LauchliA2006,SmeraldA2015} will be essential in fully pinning down the nature of these putative SN phases in \NiP{}.

\bibliographystyle{naturemag}
\bibliography{refs}

\newpage

\textbf{Methods}

{\bf Sample preparation.}
Single crystals of \NiP{} were grown by flux method~\cite{DingF2021,LiN2021}. The starting powder materials Na$_2$CO$_3$, BaCO$_3$, NiO, (NH$_4$)$_2$HPO$_4$, and NaCl were mixed thoroughly in a molar with ratio of 1:1:1:2:5. The mixture was then loaded in an alumina crucible with a lid and heated up to \qty{950}{\degreeCelsius} for \qty{20}{\hour} followed by a cooling procedure to \qty{750}{\degreeCelsius} at a slow rate of \qty{1.3}{\degreeCelsius/\hour}. The yellow single crystals of \NiP{} with layered hexagonal shape were obtained by mechanical separation from the bulk.

{\bf Magnetization and thermodynamic measurements.}
Magnetization measurements were carried out using two different magnetometers for different temperature ranges: commercial Magnetic Property Measurement System (MPMS-Quantum Design) for the temperature range $T=1.8$ -- \qty{300}{K}, and high-sensitive Hall sensor magnetometer~\cite{CandiniA2008,CavalliniA2004,CandiniA2006} for the low-temperature range $T=0.05$ -- \qty{2}{K}. The specific heat measurements were carried out using the relaxation time method, with a Quantum Design Physical Property Measurements System (PPMS) down to temperatures of \qty{0.05}{K}. The magnetocaloric effect was measured by sweeping the field up and down at \qty{25}{Oe/s} while monitoring the sample temperature with the bath temperature held fixed. The sample temperature was measured through a pre-calibrated Lakeshore Cernox semiconductor sensor~\cite{CourtsSS2003}.

{\bf Neutron scattering measurements.}
The single crystal neutron diffraction experiments were carried out using the elastic diffuse scattering spectrometer, Corelli~\cite{YeF2018}, at Spallation Neutron Source, Oak Ridge National Laboratory. 27 pieces of single crystals were co-aligned for the diffraction experiments and the total mass was about \qty{278}{mg}. The INS experiments were carried out using the time-of-flight cold neutron spectrometer, PELICAN~\cite{YuD2013}, with fixed incident energy $E_i$=\qty{3.7}{meV} at the OPAL reactor at ANSTO, and the cold neutron chopper spectrometer, CNCS~\cite{EhlersG2011}, with fixed incident energy $E_i$=1.5, \qty{3.3}{meV}. For the inelastic scattering experiments, hundreds of single crystals were co-aligned onto the oxygen-free copper sheets with total mass about 3 gram. For both spectrometers, the sample was cooled using a dilution refrigerator inserted in a \qty{7}{T} magnet with field applied along the $c$-axis of the sample. The data was processed using a combination of the freely available Dave and HORACE software~\cite{AzuahRT2009,EwingsRA2016}.

{\bf ESR measurements.}
The pulsed-field ESR measurements at $T=\qty{2}{K}$ with magnetic field up to \qty{16}{T} were conducted at the Wuhan National High Magnetic Field Center, China. The microwave frequencies in the range $140 \leq f \leq \qty{420}{GHz}$ were used in the measurements. DPPH was used as a standard in the measurements to calibrate the magnetic fields. The steady-field ESR measurements at $T=\qty{2}{K}$ with magnetic field up to \qty{22}{T} and selected microwave frequencies from \qty{58.02}{GHz} to \qty{531}{GHz} were performed at the Steady High Magnetic Field Facilities, Chinese Academy of Sciences. All the experiments were carried out using unpolarized light with InSb bolometer as the detector.

The lowest available temperature is \qty{2}{K} in our steady-field ESR experiment -- as a result, for relatively low magnetic fields ($B \gtrsim B_\text{s}$) the system could already be inside the critical fan ($T>T^*$ at fixed $B$) controlled by the QCP at $B_\text{s}$, in which region the spin correlation becomes highly short-ranged and only paramagnons survive. The lowest-energy ESR data point in Fig.~\ref{fig:2magnon}{\bf b} corresponds to $E=\qty{0.375}{meV}$ and $B=\qty{3.24}{T}$ which is already slightly inside the critical fan, and any measurement further into the critical fan becomes physically ill defined for the analysis of the 2-magnon bound state.

{\bf NMR measurements.}
The NMR measurements were conducted in a dilution refrigerator with temperature down to \qty{30}{mK}. The NMR spectra were collected by the standard spin-echo technique. The longitudinal nuclear magnetization $m(t)$ was measured by the inversion-recovery method, and the spin-lattice relaxation time was obtained by fit to the function
\begin{equation}
m(t) = m(\infty) \left[ a- 0.1e^{-(t/T_1)^\beta} - 0.9 e^{-(6t/T_1)^\beta} \right],
\end{equation}
where $\beta$ is the stretching factor. $\beta$ is about $0.7$ in the reported temperature range.

{\bf Computing the 1-magnon states.}
When the magnetic field is applied along the $c$-axis, the TL model~\eqref{eq:model_TL} is $U(1)$ symmetric. Consequently, the eigenstates can be classified by the good quantum number $\sum_i S_i^z$. Since the inter-plane exchanges are negligible compared to the in-plane coupling $J$, in this section we will only work with the TL model in one layer.

The energy of the FP state is 
\begin{equation}
E_\text{FP}=N\left( 3\Delta J -D -g_c \mu_B B \right),
\end{equation}
where $N\rightarrow \infty$ is the total number of Ni sites in each layer.

Similarly, the energy of the single-magnon excitation
$|\bm{k}\rangle \equiv \frac{1}{\sqrt{2N}}\sum_{\bm{r}}e^{\iu \bm{k}\cdot \bm{r}} S_{\bm{r}}^-|\text{FP}\rangle$ can be obtained by solving the Schr\"{o}dinger's equation, 
\begin{equation}
\left( \mathcal{H}_\text{TL} - g_c \mu_B B \sum_i S_i^z \right) |\bm{k}\rangle = \left[ E_{\text{FP}} + E_1(\bm{k}) \right]|\bm{k}\rangle,
\end{equation}
which gives
\begin{equation}
E_1(\bm{k}) = 2 \mathscr{J}(\bm{k}) - 6 \Delta J + D + g_c \mu_B B  ,
\end{equation}
and
\begin{equation}
\mathscr{J}(\bm{k}) \equiv J \left( \cos k_{x}+2\cos\frac{k_{x}}{2}\cos\frac{\sqrt{3}k_{y}}{2} \right),
\end{equation}
where we have set the in-plane lattice constant to 1.

The $T=0$ dynamic spin structure factors are given by
\begin{equation}
\mathcal{S}^{xx}(\bm{k},\omega) = \mathcal{S}^{yy}(\bm{k},\omega) = \pi \delta \left[ \omega - E_1(\bm{k}) \right],
\end{equation}
and its contribution to the INS intensity at finite-$T$ is
\begin{align}
\mathcal{I}(\bm{k},\omega) & \approx \frac{1}{1-e^{-\beta \omega}} \sum_{a,b} \left( \delta_{ab} - \hat{k}_a \hat{k}_b \right) \mathcal{S}^{ab} (\bm{k},\omega) \nonumber \\
&=\frac{\pi}{1-e^{-\beta \omega}} \delta \left[ \omega - E_1(\bm{k}) \right],\label{eq:intensity}
\end{align}
where $\beta=1/T$ is the inverse temperature. For simplicity, here we have neglected the form factors. 

{\bf Computing the 2-magnon states.}
For the 2-magnon state $|\bm{r}_1,\bm{r}_2\rangle \equiv \frac{1}{2}S_{\bm{r}_1}^- S_{\bm{r}_2}^- |\text{FP}\rangle$, the Schr\"{o}dinger's equation can be organized to the following form:
\begin{align}
\left(E-E_{\text{FP}}\right)\langle\bm{r}_{1},\bm{r}_{2}|\Psi\rangle &=\frac{1}{2}\langle\text{FP}|\left[S_{\bm{r}_{1}}^{+} S_{\bm{r}_{2}}^{+},\, \mathcal{H}_\text{TL} - g_c \mu_B B \sum_i S_i^z \right]|\Psi\rangle \nonumber \\
&= \frac{1}{2}\langle\text{FP}|S_{\bm{r}_{1}}^{+} S_{\bm{r}_{2}}^{+} \left(\mathcal{H}_0 + \mathcal{V} \right)|\Psi\rangle ,
\end{align}
where 
\begin{align}
\mathcal{H}_0 &= \sum_{i=1,2} \left(J\sum_{\bm{\eta}}\nabla_{\bm{\eta}}^{i} -6\Delta J +D+ g_c \mu_B B\right), \\
\mathcal{V} &= -J\delta_{\bm{r}_{1},\bm{r}_{2}}\sum_{\bm{\eta}}\nabla_{\bm{\eta}}^{2}+\Delta J\sum_{\bm{\eta}}\delta_{\bm{r}_{1},\bm{r}_{2}+\bm{\eta}}-2D\delta_{\bm{r}_{1},\bm{r}_{2}},
\end{align}
and the operator $\nabla_{\bm{\eta}}^i$ displaces the $i$-th magnon from $\bm{r}_i$ to $\bm{r}_i +\bm{\eta}$.

By Fourier transforming the center-of-mass coordinate to momentum space,
\begin{equation}
\Psi_{\bm{K}} (\bm{r}) = \frac{1}{\sqrt{N}} \sum_{\bm{R}} e^{-\iu \bm{K} \cdot \bm{R}} \langle \bm{R}+\frac{\bm{r}}{2},\bm{R}-\frac{\bm{r}}{2} |\Psi\rangle ,
\end{equation}
we arrive at the Lippmann-Schwinger equation:
\begin{align}
\Psi_{\bm{K}}(\bm{r}) &=\int\frac{\mathrm{d} \bm{k}}{\mathcal{A}_\text{BZ}}\frac{\cos\left(\bm{k}\cdot\bm{r}\right)}{E-E_{\text{FP}}-\left[E_{1}\left(\frac{\bm{K}}{2}+\bm{k}\right)+E_{1}\left(\frac{\bm{K}}{2}-\bm{k}\right)\right]} \nonumber \\
& \quad \cdot \left[\sum_{\bm{\eta}} J \left(- \cos\left(\frac{\bm{K}}{2}\cdot\bm{\eta}\right)+ \Delta \cos\left(\bm{k}\cdot\bm{\eta}\right)\right)\Psi_{\bm{K}}(\bm{\eta})-2D\Psi_{\bm{K}}(0)\right], \label{eq:LS}
\end{align}
where $\bm{\eta} = \{ \pm \bm{a}_1, \pm \bm{a}_2, \pm ( \bm{a}_1 + \bm{a}_2 ) \}$ are the displacement vectors of the nearest neighbor bonds, and $\mathcal{A}_\text{BZ}$ is the area of the Brillouin zone.

By setting $\bm{r}=\{0,\bm{a}_1, \bm{a}_2, \bm{a}_1+\bm{a}_2 \}$ and using the bosonic nature of the wavefunction $\Psi_{\bm{K}}(\bm{r})=\Psi_{\bm{K}}(-\bm{r})$, the Lippmann-Schwinger equation \eqref{eq:LS} is reduced to a set of coupled equations which can be solved numerically.

\textbf{Acknowledgments}\\
We would like to thank C.~D.~Batista, R.~M.~Fernandes, A.~V.~Chubukov, I.~A.~Zaliznyak and Y.~P.~Cai for helpful discussions, and L.~Sorba (NANO-CNR, Italy) for the support of the Hall sensor magnetometer substrate, X.~F.~Xiao, and J.~Y.~Zhu from Quantum Design for the support of the low temperature measurements, and G.~Davidson for the great support in setting up and operating the superconducting magnet and the dilution insert throughout the experiment on Pelican. The research was supported by the National Key Research and Development Program of China (Grant No.~2021YFA1400400, No.~2022YFA1402704, and No.~2023YFA1406500), the National Natural Science Foundation of China (Grants No.~12134020, No.~11974157, No.~12104255, No.12047501, No.~11874188, No.~12047501, No.~12005243, No.~11874080, No.~12374124, No.~12204223, No.~12474143 and No.~12374150), the Guangdong Basic and Applied Basic Research Foundation (Grant No.~2021B1515120015, No.~2022B1515120014, No.~2023B0303000003 and No.~2023B1515120060), the Shenzhen Fundamental Research Program (Grant No. JCYJ20220818100405013 and JCYJ20230807093204010).The Major Science and Technology Infrastructure Project of Material Genome Big-science Facilities Platform was supported by Municipal Development and Reform Commission of Shenzhen. A portion of this research used resources at the Spallation Neutron Source, a DOE Office of Science User Facility operated by the Oak Ridge National Laboratory. Z.W.'s theoretical calculations at University of Minnesota were supported by the U.S. Department of Energy through the University of Minnesota Center for Quantum Materials under Award No. DE-SC-0016371. During the writing of the paper, 
Z.W. was supported by the National Key Research and Development Program of China (Grant No. 2022YFA1402200), and the Key Research and Development Program of Zhejiang Province, China (Grant No. 2021C01002). The authors would also like to acknowledge the beam time awarded by ANSTO through the proposal No.~P9955 and SNS ORNL through the proposal No.~29333.

\textbf{Author contributions}\\
J.M.S., Z.W., L.S.W., J.W.M., D.H.Y., W.Y. and D.P.Y. designed the experiments. W.R.J., L.W., J.Y. and Y.F. provided single crystals used in this study. J.M.S., H.G., N.Z., T.T.L., S.M.W., J.L.Z., X.T. and L.S.W. carried out the low temperature measurements. F.Y., J.M.S., P.M., A.P., L.S.W., R.M. and D.H.Y. carried out the neutron scattering experiments. J.M.S, Z.W.O., W.T., Z.T.Z. and L.M. carried out the ESR experiments. 
X.X., Y.C., and W.Y. performed the NMR measurements.
A.C. and G.B. designed and fabricated the hall sensor magnetometer used in this study. Z.W., L.W., L.X., J.Z.Z., B.X. and J.W.M. developed the theoretical explanations. Z.W. proposed and carried out the 1- to 4-magnon calculations and the perturbative calculations. L.W. and L.X. performed the DMRG calculations. All authors discussed the results and contributed to the writing of the manuscript.

\textbf{Competing interests}\\
The authors declare no competing interests.

\newpage

\setcounter{figure}{0} 
\setcounter{equation}{0} 
\renewcommand{\thefigure}{S\arabic{figure}}
\renewcommand{\thetable}{S\arabic{table}}
\renewcommand{\theequation}{S\arabic{equation}}

\begin{center}   
{\bf Supplementary Information: ``Bose-Einstein condensation of a two-magnon bound state in a spin-one triangular lattice''} 
\end{center}

\textbf{Table of Contents }

Supplementary Note 1: Single crystal X-ray diffraction refinements\par
Supplementary Note 2: Magnetization measurements\par
Supplementary Note 3: Specific heat measurements\par
Supplementary Note 4: Magnetic structure analysis\par
Supplementary Note 5: Scaling analysis of the phase boundary\par
Supplementary Note 6: Neutron scattering measurements\par
Supplementary Note 7: Magnetocaloric Effect (MCE)\par
Supplementary Note 8: ESR spectra\par  
Supplementary Note 9: NMR measurements\par  
Supplementary Note 10: 3- and 4-magnon states\par   
Supplementary Note 11: DMRG calculation\par
Table S1. Single crystal X-ray diffraction refinements for \NiP \par
Table S2. Wyckoff positions, coordinates, occupancies, and equivalent isotropic displacement \par
parameters for \NiP \par
Table S3. Basis vectors of decomposed irreducible representations \par
Figure S1. Temperature dependent magnetization of \NiP{} with field along the $\bm{c}$ axis \par
Figure S2. Temperature dependent magnetization of \NiP{} with field in the $\bm{ab}$ plane \par
Figure S3. Field dependence of magnetization $M(B)$ measured at different temperatures with \par
field applied along the $\bm{c}$ and $\bm{a}$ axes \par
Figure S4. Temperature- and field-evolution of magnetic specific heat with $B \parallel \bm{c}$ \par
Figure S5. Temperature- and field-evolution of magnetic specific heat with $B \parallel \bm{a}$ \par
Figure S6. Field-temperature magnetic phase diagram overlaid on contour plots of the \par 
normalized entropy $S/R \ln 3$, with $B \parallel \bm{c}$ and $B \parallel \bm{a}$ \par
Figure S7. Temperature dependence of magnetic specific heat and magnetic entropy at $B=\qty{0.8}{T}$ \par with $B \parallel \bm{c}$ \par
Figure S8. Neutron diffraction data measured in the $(H,K,L=0)$ and $(H,K=0,L)$ scattering \par
plane at $T=\qty{80}{mK}$ and $B=\qty{0}{T}$\par
Figure S9. Possible zero-field magnetic structures and calculated magnetic structure factor \par
versus observation for three different IRs $\Gamma_1$, $\Gamma_2$ and $\Gamma_3$ \par
Figure S10. Scaling behavior of the magnetic phase boundary $T\propto(B_{\rm s}-B)^{\nu z}$ \par
Figure S11. Order parameter of the magnetic reflections \par
Figure S12. Inelastic neutron scattering intensity measured at $T=\qty{60}{mK}$ and $B=\qty{3}{T}$ \par
Figure S13. Inelastic neutron scattering spectra along high-symmetry directions measured at \par
$T=\qty{60}{mK}$ and different magnetic fields \par
Figure S14. Magnetization and MCE measured at \qty{50}{mK} with the field applied along the $\bm{c}$ axis  \par
Figure S15. Magnetic susceptibility $dM/dB$ and MCE measured at different temperatures with \par 
magnetic field along the $\bm{c}$ axis \par
Figure S16. ESR spectra measured at $T=\qty{2}{K}$ in a pulsed magnetic field for $B \parallel \bm{ab}$ and $B \parallel \bm{c}$ \par
Figure S17. ESR spectra measured at $T=\qty{2}{K}$ in a steady magnetic field tilted $6^\circ$ away from the \par $\bm{c}$ axis \par
Figure S18. ESR spectra measured at $T=\qty{2}{K}$ with frequencies $F=\qty{58.02}{GHz}$$\qty{64.02}{GHz}$ in a \par steady magnetic field tilted $6^\circ$ away from the $\bm{c}$ axis \par 
Figure S19. The calculated frequency-field relation of the 1-magnon state and the 2-magnon \par bound state at the $\Gamma$ point \par
Figure S20. Spin-lattice relaxation rate $1/T_1$ measured at different fixed fields with $B \parallel \bm{c}$ and \par the energy-field relation of 2-magnon bound state observed in NMR measurements  \par
Figure S21. Initial basis for 2-, 3-, and 4-magnon states used in the variational method \par 
Figure S22. Results of the variational calculation for $\mathcal{H}_\text{TL}$ \par 
Figure S23. Field dependence of the $T=0$ magnetization for cylinders of size {\bf a} $9\times 6$, {\bf b} $12\times 6$, \par and {\bf c} $15\times 6$  \par 
Figure S24. Field dependence of the $T=0$ magnetization for cylinders of size {\bf a} $6\times 4$, {\bf b} $9\times 6$, \par and {\bf c} $12\times 8$ \par
Figure S25. Extrapolation of critical fields $\widetilde{H}_{\rm c1}$, $\widetilde{H}_{\rm c2}$, and $\widetilde{H}_{\rm s}$ as a function of $1/L_y$ to the \par thermodynamic limit \par 
Figure S26. Dipolar and quadrupolar structure factors at $\Gamma$ and K points \par

\newpage

\section{Single crystal X-ray diffraction refinements }
\begin{table}[bp!]
\caption{Single crystal X-ray diffraction refinements for \NiP.}
\begin{ruledtabular}
\begin{tabular}{lc}
\textbf{Formula}                  & \textbf{Na$_2$BaNi(PO$_4$)$_2$} \\
\hline
     Formula mass (g/mol)             &       431.95                   \\
     Crystal system                   &       Trigonal                  \\
     Space group                      &       $P\bar{3}m1$                  \\
     $a$(\AA{})                      &       5.2806(4)                 \\
     $b$({\AA})                      &       5.2806(4)                 \\
     $c$({\AA})                      &       6.9591(5)                 \\
     $\alpha$                         &       $90^\circ$                    \\
     $\beta$                          &       $90^\circ$                    \\
     $\gamma$                         &       $120^\circ$                    \\
     $V (\text{\AA}^3)$                    &       168.054(5)                \\
     $T (K)$                          &       100                       \\
     $\rho(cal)(g/cm^3)$              &       4.269                     \\
     $\lambda $({\AA})                &       0.71073                   \\
      F (000)                         &       214.0                      \\
     Crystal size (mm$^3$)            &       $0.11\times0.06\times0.02$            \\
     $\mu$ (mm$^{-1}$)                &       17.454                       \\
     Final R indices                  &       R$_1$ = 3.83, $\omega$R$_2$ = 5.95    \\
     R indices (all data)             &       R$_1$ = 3.98, $\omega$R$_2$ = 6.00    \\
     Goodness of fit                  &       2.97                      \\
\end{tabular}
\end{ruledtabular}\label{structure}
\end{table}

\begin{table}[tbp!]
\caption{Wyckoff positions, coordinates, occupancies, and equivalent isotropic displacement parameters for \NiP.}
\begin{ruledtabular}
\begin{tabular}{lcccccc}
\textbf{Atom}  & \textbf{Wyckoff site} & \textbf{$x$} & \textbf{$y$} & \textbf{$z$} & \textbf{Occupancy} & \textbf{$\rm U_{eq}$}\\
\hline
Ba$_1$        &       1$a$            &   0          &   0          &       0      &     1              &   0.00310   \\
Ni$_1$        &       1$b$            &   0          &   0          &      0.5     &     1              &   0.00593   \\
Na$_1$        &       2$d$            &   0.666667   &   0.333333   &   0.681298   &     1              &   0.01090   \\
P$_1$        &       2$d$            &   0.666667   &   0.333333   &   0.243700   &     1              &   0.00556   \\
O$_1$        &       2$d$            &   0.666667   &   0.333333   &   0.026836   &     1              &   0.00814   \\
O$_2$        &       6$i$            &   0.356792   &   0.178396   &   0.323064   &     1              &   0.03942   \\
\end{tabular}
\end{ruledtabular}\label{structure1}
\end{table}

High quality single crystals of \NiP{} were grown using the flux method, and the crystal structure of \NiP{} was characterized using a Bruker D8 Quest diffractometer with Mo-K$\alpha$ radiation ($\lambda=\qty{0.71073}{\AA{}}$). The data integration and reduction were performed with the commercial Bruker APEX2 software suite. The refined lattice parameters and the atomic occupations are presented in Tables~\ref{structure} and \ref{structure1}. We note that a slightly different crystal structure with space group $P\bar{3}$ was reported~\cite{LiN2021,DingF2021}. We also tried to refine the crystal structure with $P\bar{3}$, but obtained higher R-factors: R$_1$ = 6.57, $\omega$R$_2$ = 8.90.
\newpage
~
\newpage

\section{Magnetization measurements}
\begin{figure}[ht!]
  \centering
  \includegraphics[width=0.95\columnwidth]{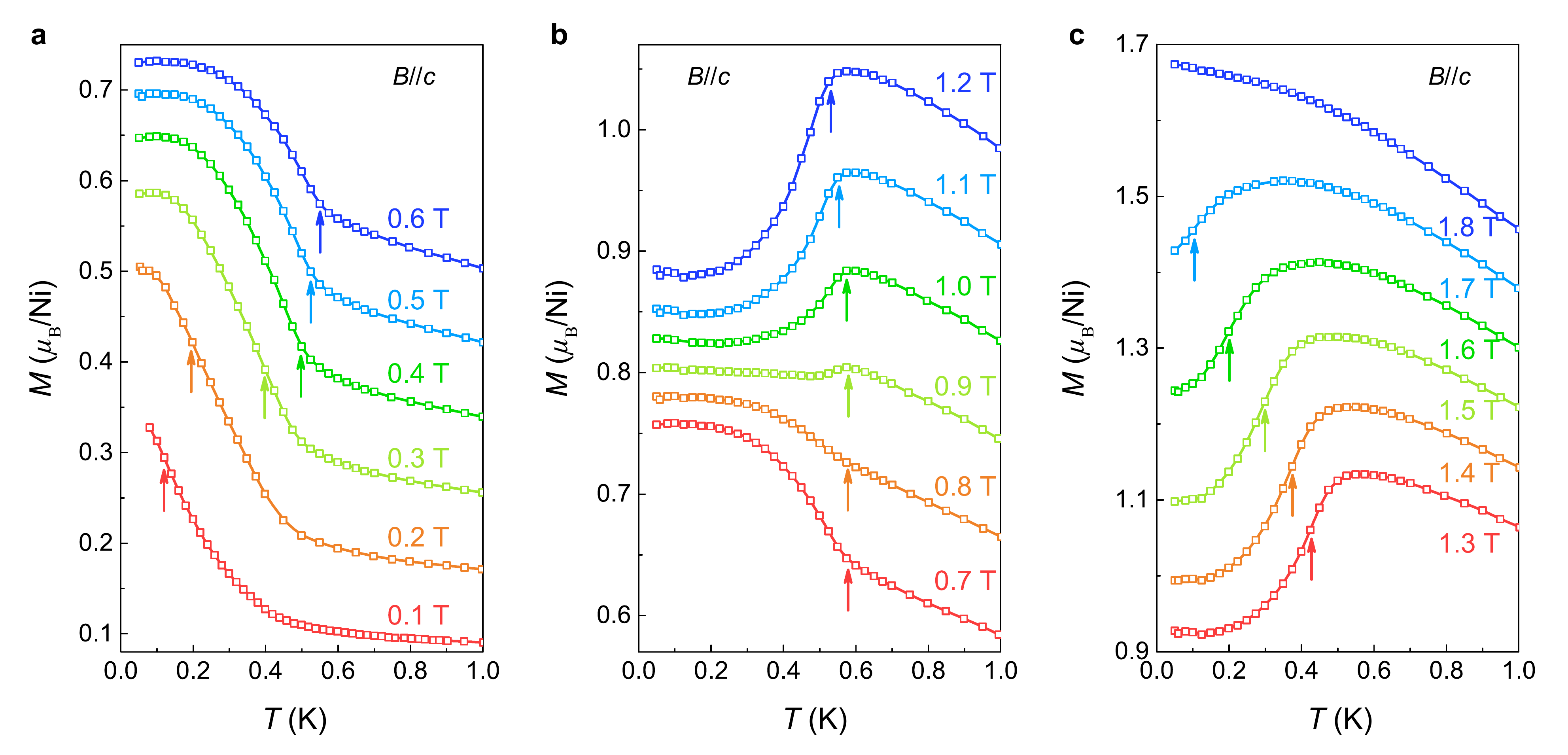}
  \caption{Temperature dependent magnetization of \NiP{} with field along the $\bm{c}$ axis. The arrows indicate the transition temperatures at different fields.}
  \label{MTc}
\end{figure}

\begin{figure}[ht!]
  \centering
  \includegraphics[width=0.95\columnwidth]{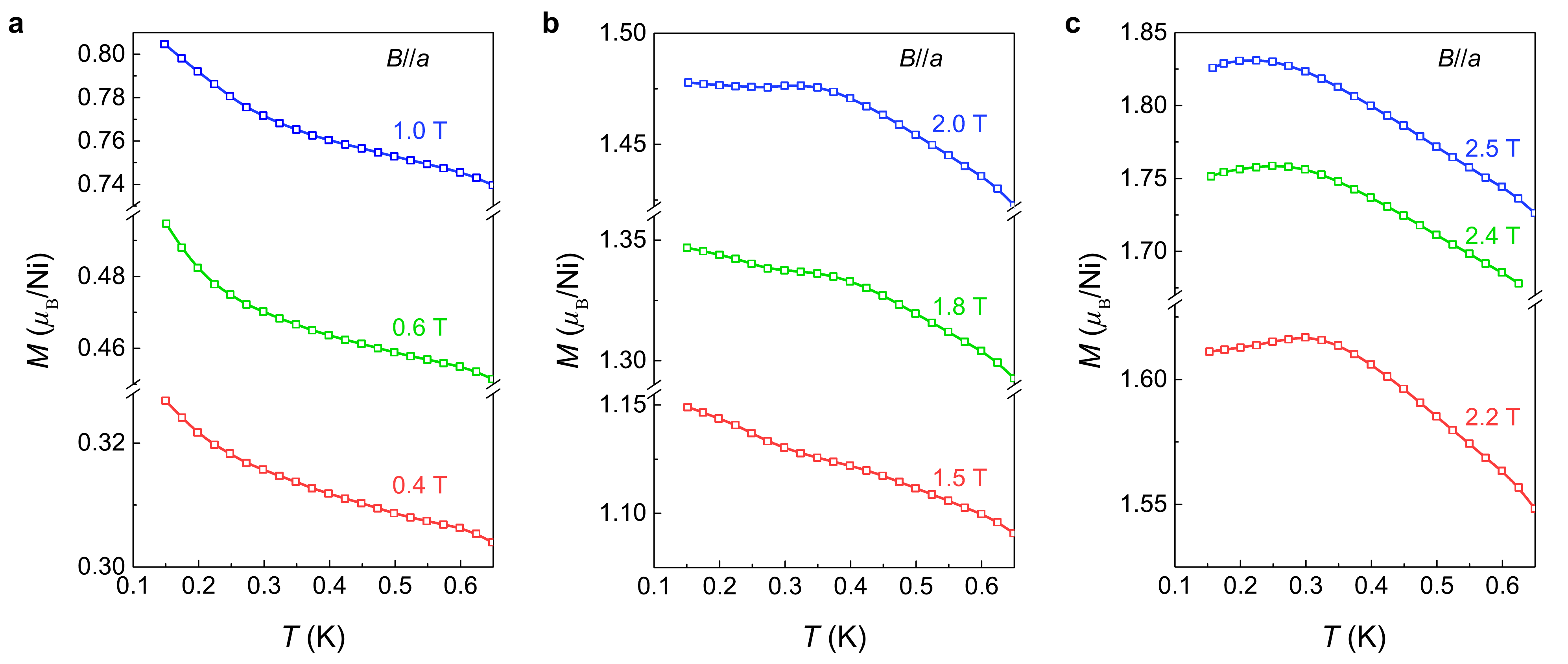}
  \caption{Temperature dependent magnetization of \NiP{} with field in the $\bm{ab}$ plane.}
  \label{MTa}
\end{figure}

The temperature dependent magnetization of \NiP{} with field along the $\bm{c}$ and $\bm{a}$ axes are shown in Figs.~\ref{MTc} and \ref{MTa}, respectively. 
All the transition temperatures were determined through the peak positions of the temperature derivative $dM/dT$, indicated by the arrows shown in Fig.~\ref{MTc}.

\begin{figure}[ht!]
\centering
\includegraphics[width=0.99\columnwidth]{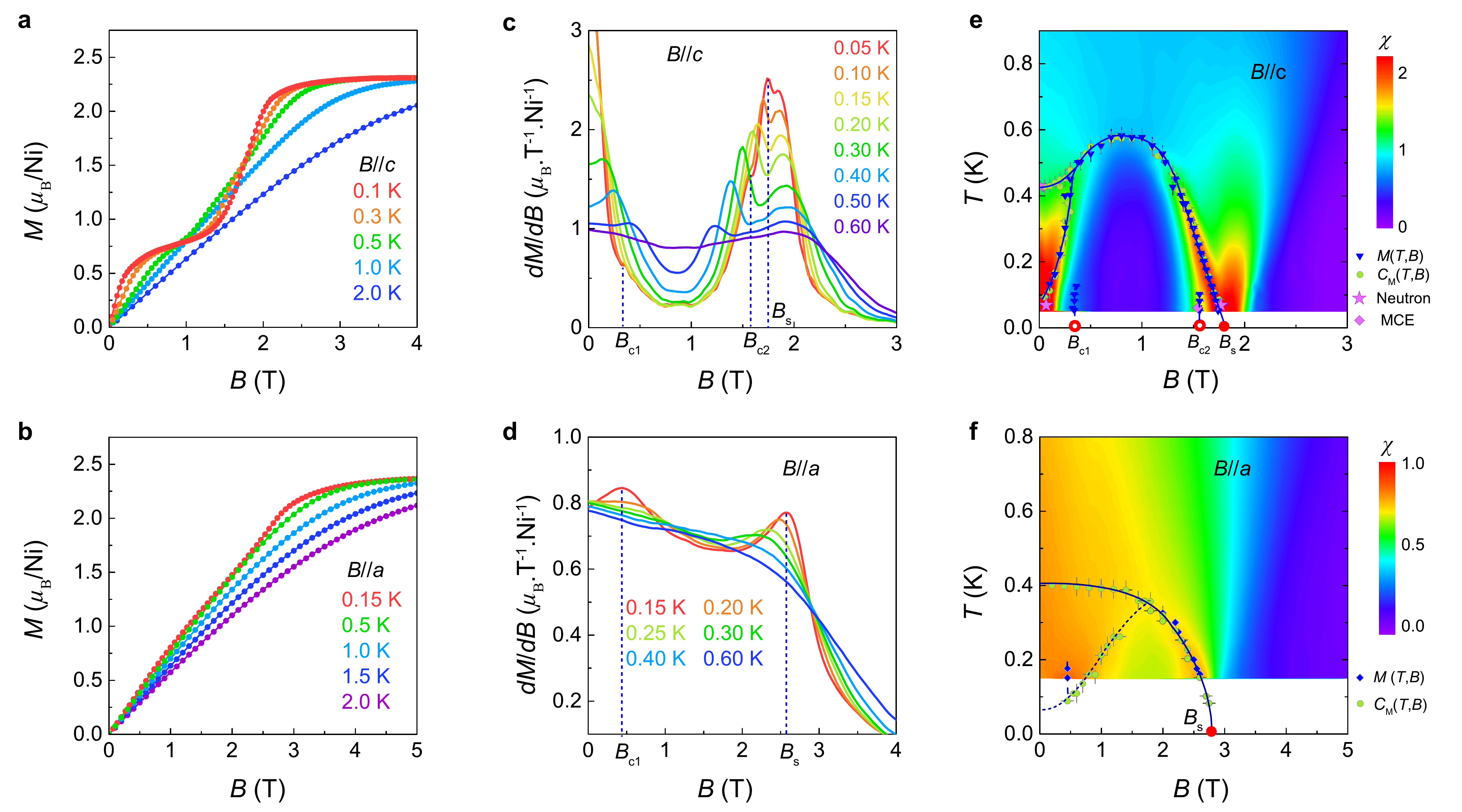}
\caption{{\bf a}-{\bf b} Field dependence of magnetization $M(B)$ measured at different temperatures with field applied along the $\bm{c}$ and $\bm{a}$ axes, respectively. {\bf c}-{\bf d} Magnetic susceptibilities $dM/dB$ measured at different temperatures with field applied along the $\bm{c}$ and $\bm{a}$ axes, respectively. {\bf e}-{\bf f} The field-temperature magnetic phase diagram overlaid on the contour plots of magnetic susceptibility $dM/dB$ with magnetic field along the $\bm{c}$ and $\bm{a}$ axes, respectively. The magnetic phase boundaries were extracted through $M(T,B)$, $C(T,B)$, neutron diffraction and MCE measurements. Error bars indicate standard deviations.}
\label{MB}
\end{figure}

Figures~\ref{MB}{\bf a}-{\bf b} show the isothermal magnetization $M(B)$ measured at various temperatures with $B \parallel \bm{c}$ and $B \parallel \bm{a}$, respectively. As we lower the temperature, the phase boundaries gradually become pronounced for both field directions. For $B \parallel \bm{c}$, a broad plateau is revealed at about one third of the saturated moment for $T\lesssim \qty{0.3}{K}$. In contrast, no plateau phase was observed for the field in the $\bm{ab}$ plane. The temperature evolution of the anomalies in the $dM/dB$ curves for both field directions is illustrated in Figs.~\ref{MB}{\bf c}-{\bf d}. 
The contour plots of the magnetic susceptibility $dM/dB$ as a function of temperature and field for both field directions are presented in Figs.~\ref{MB}{\bf e}-{\bf f}. The magnetic phase boundaries extracted through $M(T,B)$, $C(T,B)$, neutron diffraction, and magnetocaloric effect (MCE) measurements are over-plotted.

\newpage

\section{Specific heat measurements}
\begin{figure}[bp!]
\centering
\includegraphics[width=0.85\columnwidth]{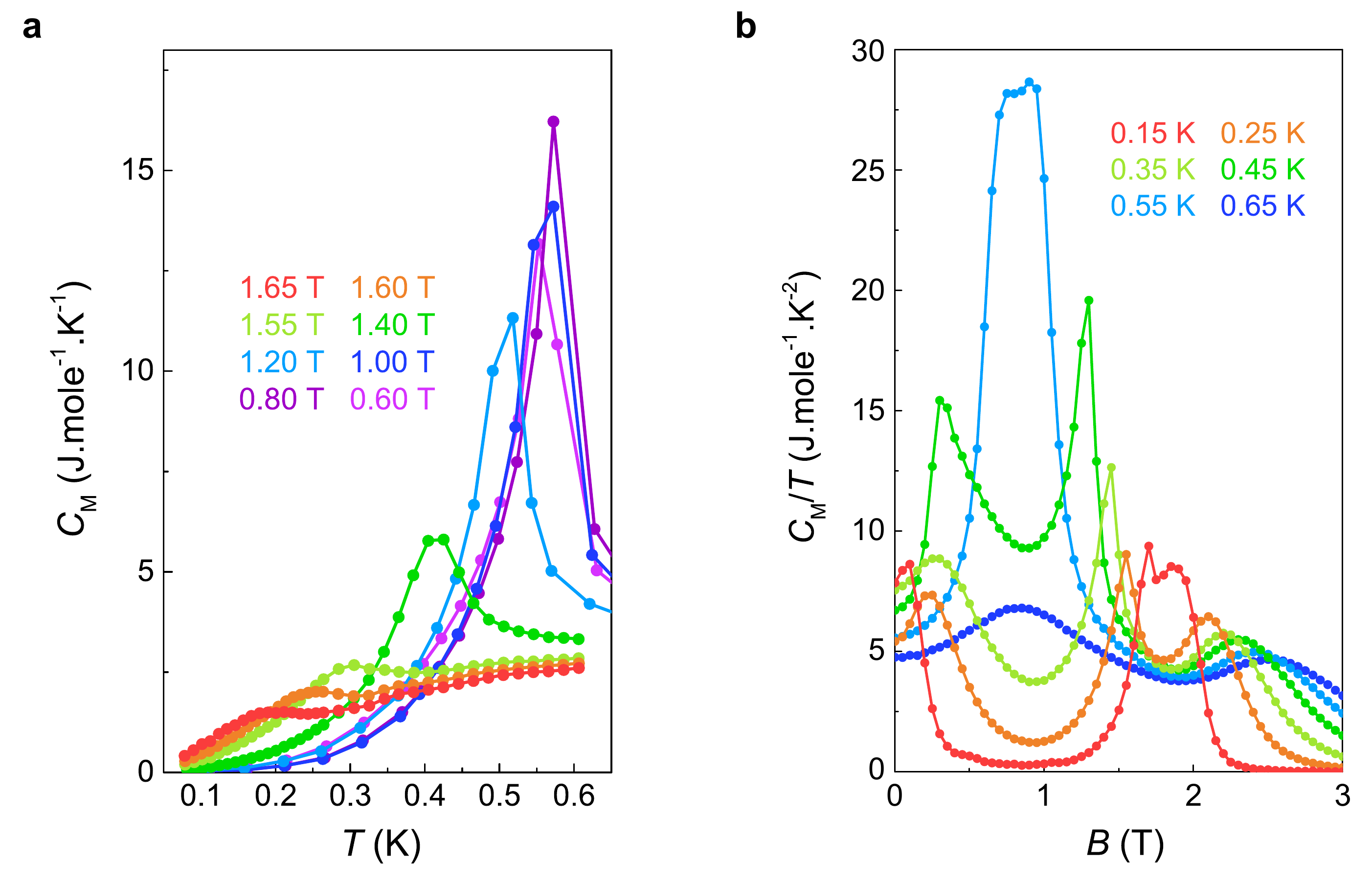}
\caption{{\bf a} Temperature evolution of the magnetic specific heat $C_{\rm M}$ measured at different magnetic fields with $B \parallel \bm{c}$. {\bf b} Field evolution of magnetic specific heat $C_{\rm M}/T$ measured at different temperatures with $B \parallel \bm{c}$. }
\label{HC_c}
\end{figure}

\begin{figure}[tbp!]
\centering
\includegraphics[width=0.85\columnwidth]{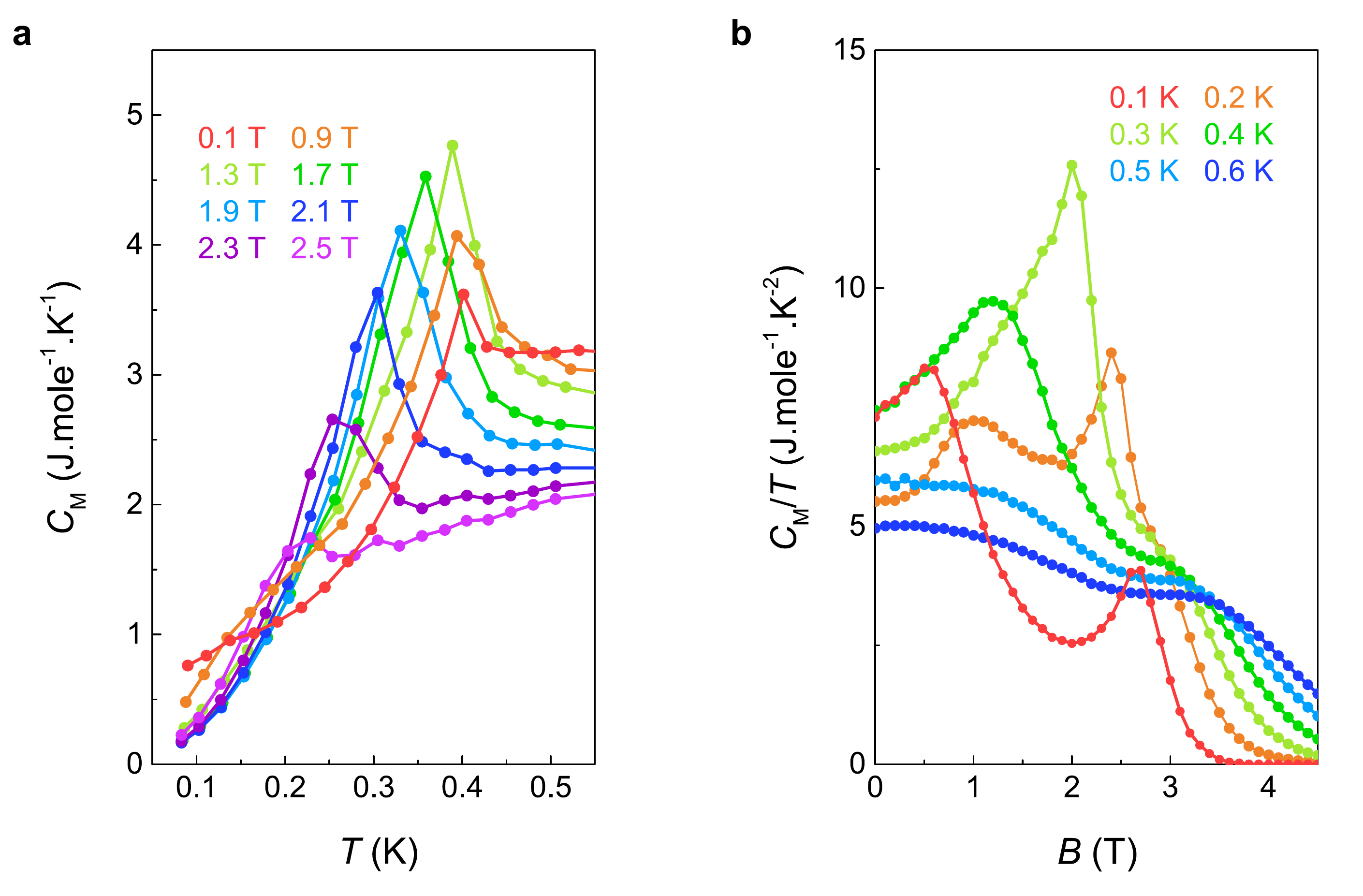}
\caption{{\bf a} Temperature evolution of the magnetic specific heat $C_{\rm M}$ measured at different magnetic fields with $B \parallel \bm{a}$. {\bf b} Field evolution of the magnetic specific heat $C_{\rm M}/T$ measured at different temperatures with $B \parallel \bm{a}$. }
\label{HC_a}
\end{figure}

The low-temperature specific heat of \NiP{} with magnetic field applied along the $\bm{c}$ and $\bm{a}$ axes are shown in Figs.~\ref{HC_c}-\ref{HC_a}, respectively. For both field directions, clear sharp anomalies were observed and the peak positions were extracted as the upper phase boundaries, shown as circles in Fig.~2{\bf e} in the main text and Figs.~\ref{MB}{\bf e}-{\bf f} in Supplementary Information.

\begin{figure}[ht!]
\centering
\includegraphics[width=0.95\columnwidth]{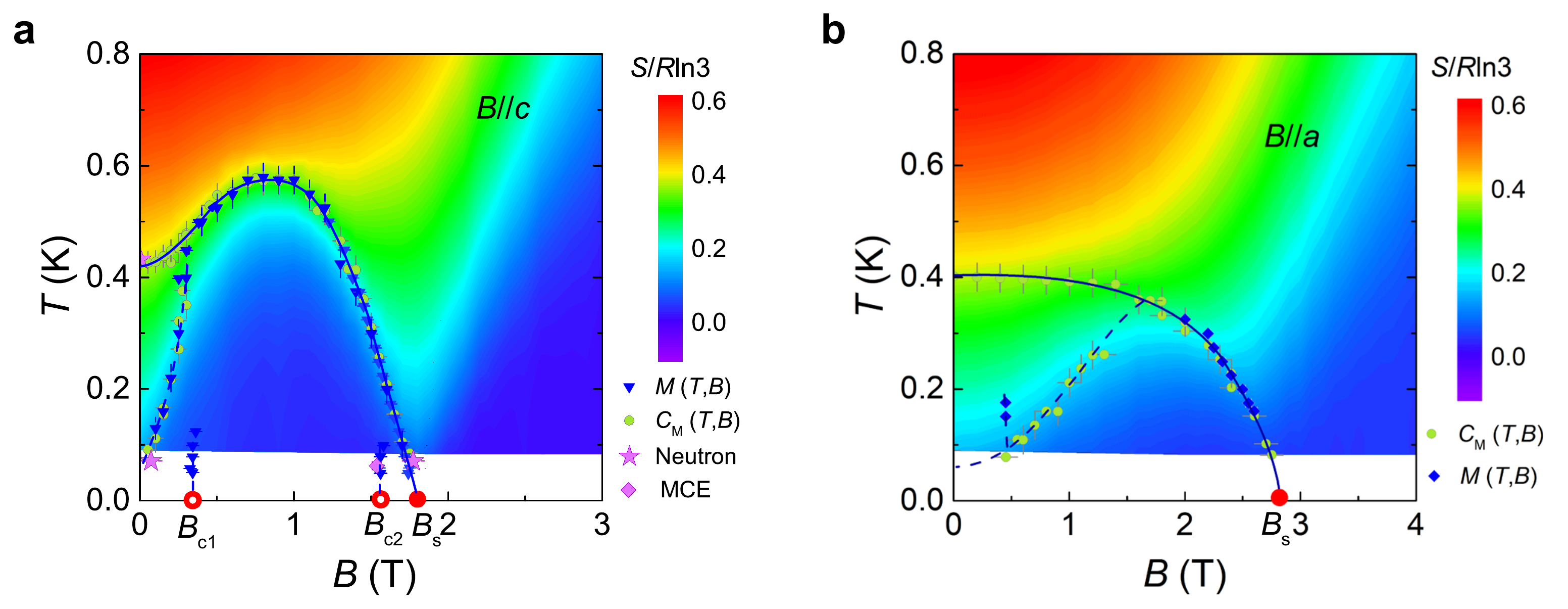}
\caption{Field-temperature magnetic phase diagram overlaid on contour plots of the normalized entropy $S/R \ln 3$, with $B \parallel \bm{c}$ and $B \parallel \bm{a}$, respectively. The magnetic phase boundaries were extracted through $M(T,B)$, $C(T,B)$, neutron diffraction and MCE measurements. Error bars indicate standard deviations}
\label{entropy}
\end{figure}

The magnetic phase diagrams overlaid on contour plots of normalized entropy $S/R \ln 3$ for both field directions are shown in Fig.~\ref{entropy}. It is clear that only a small portion of the full $R\ln 3$ entropy is released (about 0.2$R \ln3$ to 0.4$R \ln 3$) at the upper phase boundary. In other words, a great amount of moments are still fluctuating deep inside these ordered phases.

\begin{figure*}[h]
  \centering
  \includegraphics[width=0.45\columnwidth]{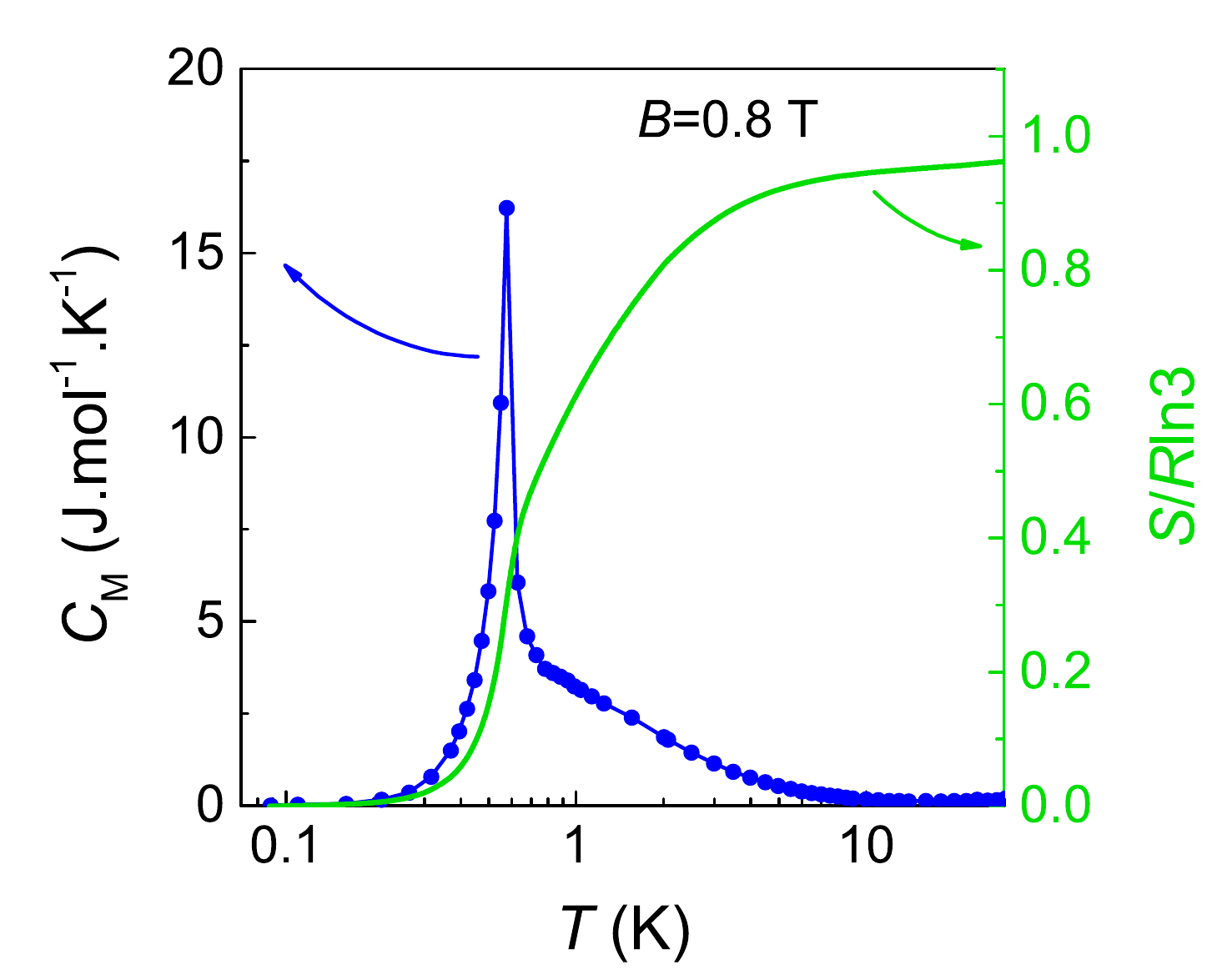}
  \caption{Left axis: Temperature dependence of the magnetic specific heat (blue filled circles) measured at $B=\qty{0.8}{T}$ with $B \parallel \bm{c}$.  Right axis: The temperature dependence of the integrated magnetic entropy at $B=\qty{0.8}{T}$ (green solid line).}
  \label{B_0p8T_entropy}
\end{figure*}

Since spin fluctuations persist down to the inaccessible low-temperature region,  we have extrapolated a residual entropy of $0.15 R\ln 3$ at $B=0$ in the main text. Note that the spin fluctuations are greatly reduced in the $1/3$-plateau (see Fig.~\ref{entropy}{\bf a}), as a result the full $R \ln 3$ is expected to be recovered in this phase without accounting for the inaccessible low-temperature region.
Figure~\ref{B_0p8T_entropy} shows the temperature-dependent magnetic specific heat (blue filled circles) measured at $B=\qty{0.8}{T}$. The corresponding integrated entropy (green soild line) is also presented, without adding a residual entropy by hand. We can find that the total entropy reaches $0.96 R\ln 3$ above \qty{10}{K}, clearly demonstrating the effective spin-1 physics for \NiP.

\newpage
~ 
\newpage

\section{Magnetic structure analysis}

The magnetic structure of \NiP{} at \qty{80}{mK} and $B=0$ has four inequivalent propagation vectors: $\bm{k}_1= (1/3,1/3,0.293)$, $\bm{k}_2= (1/3,1/3,-0.293)$, $\bm{k}_3= (-1/3,-1/3,0.293)$, and $\bm{k}_4= (-1/3,-1/3,-0.293)$, which are not connected by the primitive vectors in reciprocal space. This set of propagation vectors allows for either multi-$k$ ordering or single-$k$ ordering with four equally weighted magnetic domains.

\begin{figure}[bp!]
\centering\includegraphics[width=0.99\columnwidth]{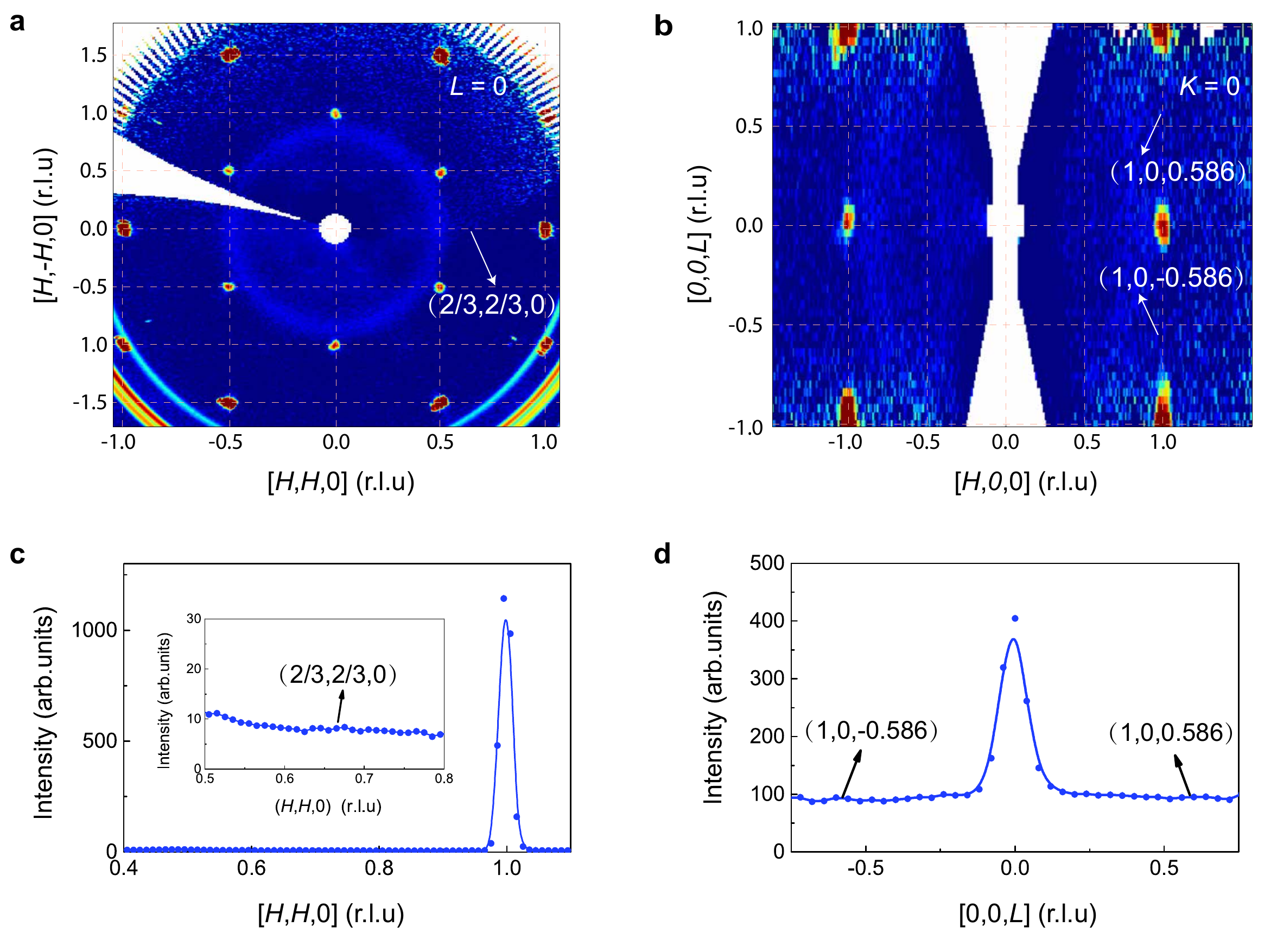}
\caption{Neutron diffraction pattern measured in the {\bf a} $(H,K,L=0)$ and {\bf b} $(H,K=0,L)$ scattering plane at $T=\qty{80}{mK}$ (slightly above $T_{\rm N1}$) and $B=\qty{0}{T}$. The line cuts along the {\bf c} [H,H,0] and {\bf d} [0,0,L] directions, respectively. There is no extra higher harmonic magnetic Bragg peaks observed around (2/3,2/3,0) and (1,0,$\pm0.586$) positions.}
\label{High-harmonics}
\end{figure}

In the case of multi-$k$ ordering, magnetic Bragg peaks are typically anticipated at higher harmonics, such as ($\pm 2/3$,$\pm 2/3$,0), (0,0,$\pm 0.586$), and (0,0,0), resulting from different linear combinations of the propagation vectors. 
Figure~\ref{High-harmonics} shows a close examination of the higher harmonic positions, where no visible peak could be observed at ($\pm 2/3$,$\pm 2/3$,0) or (1,0,$\pm 0.586$). 
In other words, there is unlikely any formation of mutli-$k$ structure which involves both $\bm{k}_i$ and $\bm{k}_j$ where $i \in \{1,4\}$ and $j \in \{2,3\}$. 
Another possibility is the double-$k$ structure with $\bm{k}_i$ and $\bm{k}_j \equiv -\bm{k}_i$ (e.g., $\bm{k}_1$ and $\bm{k}_4$), where  
the higher harmonic positions overlap with the nuclear peaks, which makes determining the existence of higher-harmonic peak a highly challenging task. In the following, we assumed single-$k$ structures with multiple domains to analyze the neutron diffraction data.

\begin{table}[tbp!]
  \caption{Basis vectors of decomposed irreducible representations (IR) of space group $P\bar{3}m1$ with magnetic wave-vector $ \bm{k}_1 =(1/3,1/3,0.293)$ and moments on Ni site (0,0,0.5).}
\begin{ruledtabular}
\begin{tabular}{lcccccc}
     \textbf{IR}  & \textbf{$m_a$} & \textbf{$m_b$} & \textbf{$m_c$} & \textbf{$im_a$} & \textbf{$im_b$} & \textbf{$im_c$}\\
                                                             \hline
     $\Gamma_1$        &       0        &   0            &   1            &       0         &     0           &   0\\
     $\Gamma_2$        &       1        &   0            &   0            &    $1/\sqrt{3}$       &   $2/\sqrt{3}$        &   0\\
     $\Gamma_3$        &       1        &   0             &   0           &   $-1/\sqrt{3}$       &  $-2/\sqrt{3}$        &   0\\
\end{tabular}
\end{ruledtabular}\label{BV}
\end{table}

\begin{figure}[tbp!]
\centering\includegraphics[width=0.99\columnwidth]{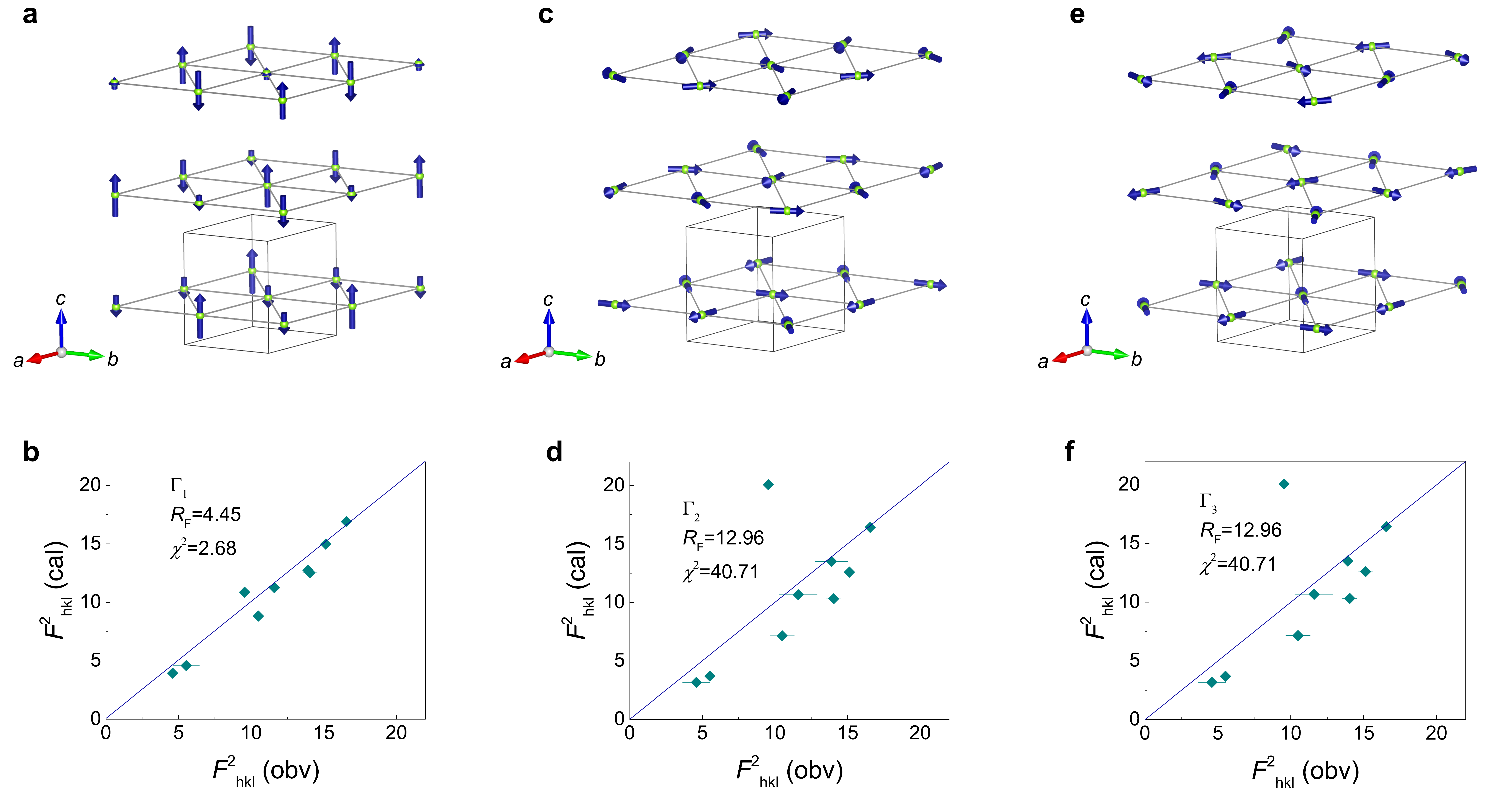}
\caption{{\bf a}, {\bf c} and {\bf e} Possible zero-field magnetic structures of \NiP{} correspond to three different IRs $\Gamma_1$, $\Gamma_2$ and $\Gamma_3$, respectively. {\bf b}, {\bf d} and {\bf f} The calculated magnetic structure factor versus observation using $\Gamma_1$, $\Gamma_2$ and $\Gamma_3$ model, respectively. Error bars indicate standard deviations.}
  \label{Mag_str}
\end{figure}

The magnetic structure at \qty{80}{mK} and $B=0$ is analyzed by representation theory. For the space group $P\bar{3}m1$ with Ni site at (0,0,0.5) and the magnetic propagation vector \{$\bm{k}_1$, $\bm{k}_2$, $\bm{k}_3$, $\bm{k}_4$\}, the spin configuration can be described by three different irreducible representations (IRs). Table~\ref{BV} lists the basis vectors for IRs $\Gamma_1$, $\Gamma_2$ and $\Gamma_3$, respectively.

For $\Gamma_1$, the moments are constrained along the $\bm{c}$ axis with an ``UUD'' spin configuration in the $\bm{ab}$ plane (Fig.~\ref{Mag_str}{\bf a}). The moment on each Ni site is modulated by $M{_A}e^{ \iu k_z \cdot r_z}$ along the $\bm{c}$-axis, where $M{_A}$ is the amplitude of Ni$^{2+}$ moment, $k_z=0.293$, $r_z$ is the $z$ component of Ni$^{2+}$ coordinate. For $\Gamma_2$ and $\Gamma_3$, the magnetic structures are degenerate $120^\circ$ spin structures with anti-phase chirality. The moments rotate along the $\bm{c}$-axis with an angle $\cos^{-1}(k_z \cdot r_z)$, as shown in Figs.~\ref{Mag_str}{\bf c} and {\bf e}, respectively. 



The software FullProf Suite was applied to determine the magnetic structure of \NiP{}. To enhance statistics, we averaged the intensities of equivalent magnetic reflections from the four magnetic domains. Magnetic structure refinements were performed by considering each of the irreducible representations ($\Gamma_1$, $\Gamma_2$, $\Gamma_3$) for the $\bm{k}_1$= (1/3,1/3,0.293) propagation vector. We found that $\Gamma_1$ gives the best goodness of fit with $R_F=4.45$, $\chi^2=2.68$ for the observed data, as shown in Figs.~\ref{Mag_str}{\bf b}, {\bf d} and {\bf f}. Additionally, potential spin configurations involving linear combinations of $\Gamma_1$, $\Gamma_2$, and $\Gamma_3$ were also considered. However, the refinement consistently converged to the spin configuration of $\Gamma_1$, with the ratio of $\Gamma_2/\Gamma_1$ or $\Gamma_3/\Gamma_1$ approaching zero. All refinement results clearly indicate that the magnetic structure of the system is collinear, featuring an ``UUD'' spin configuration in the $\bm{ab}$ plane, illustrated in Fig.3{\bf b} of the main text. The amplitude of the moment determined from the refinement is $M_A=\qty{1.40(1)}{\mu_B}$.  

\newpage
\section{Scaling analysis of the phase boundary}
\begin{figure}[h]
\centering
\includegraphics[width=0.9\columnwidth]{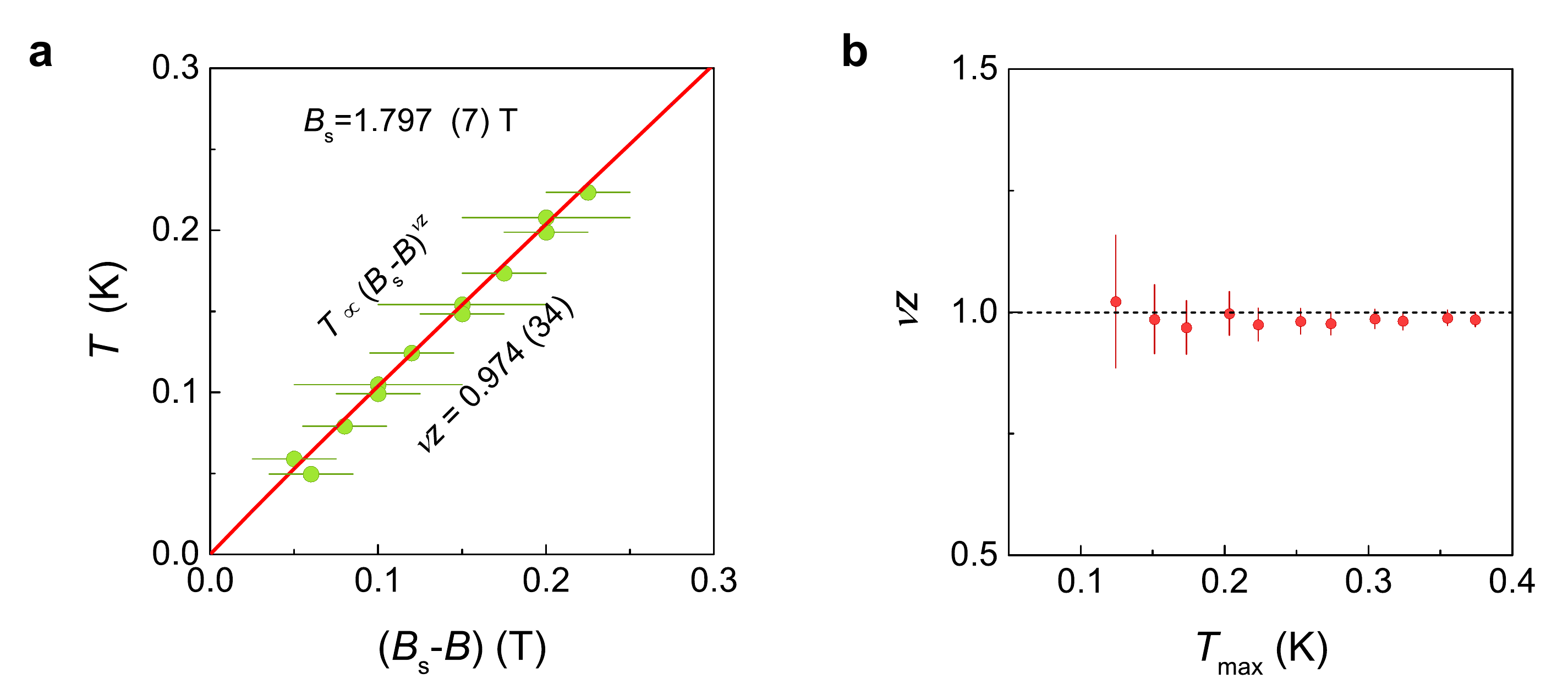}
\caption{ {\bf a} Scaling behavior of the magnetic phase boundary $T\propto(B_{\rm s}-B)^{\nu z}$ with the highest temperature of fitting range $T_\text{max}$=\qty{223.4}{mK} $\lesssim 0.4T_{c \text{ max}}$ in the vicinity the critical field $B_{\rm s}$. {\bf b} Critical exponents ${\nu z}$ as a function of $T_\text{max}$, with fitting performed for \qty{49.5}{mK} $\leq T \leq T_\text{max}$. Error bars indicate standard deviations in {\bf a} and {\bf b}.}
\label{scaling}
\end{figure}

The scaling analysis of the exponents should be performed in a low-temperature window~\cite{ZapfV2014_RMP,TanakaH2007}. To verify the exponents extracted in this work are robust and valid, we tried to fit them for a series of window sizes to check convergence. In Fig.~\ref{scaling}, we show that $\nu z$ stays practically at 1 across a wide range of fitting windows: from $T_\text{max} \approx \qty{374.1}{mK}\sim 0.65T_\text{c max}$ down to $T_\text{max} \approx \qty{124.2}{mK}$, confirming that the fitting is robust in a wide critical region.

\newpage

\section{Neutron scattering}

\begin{figure}[bp!]
\centering
\includegraphics[width=0.85\columnwidth]{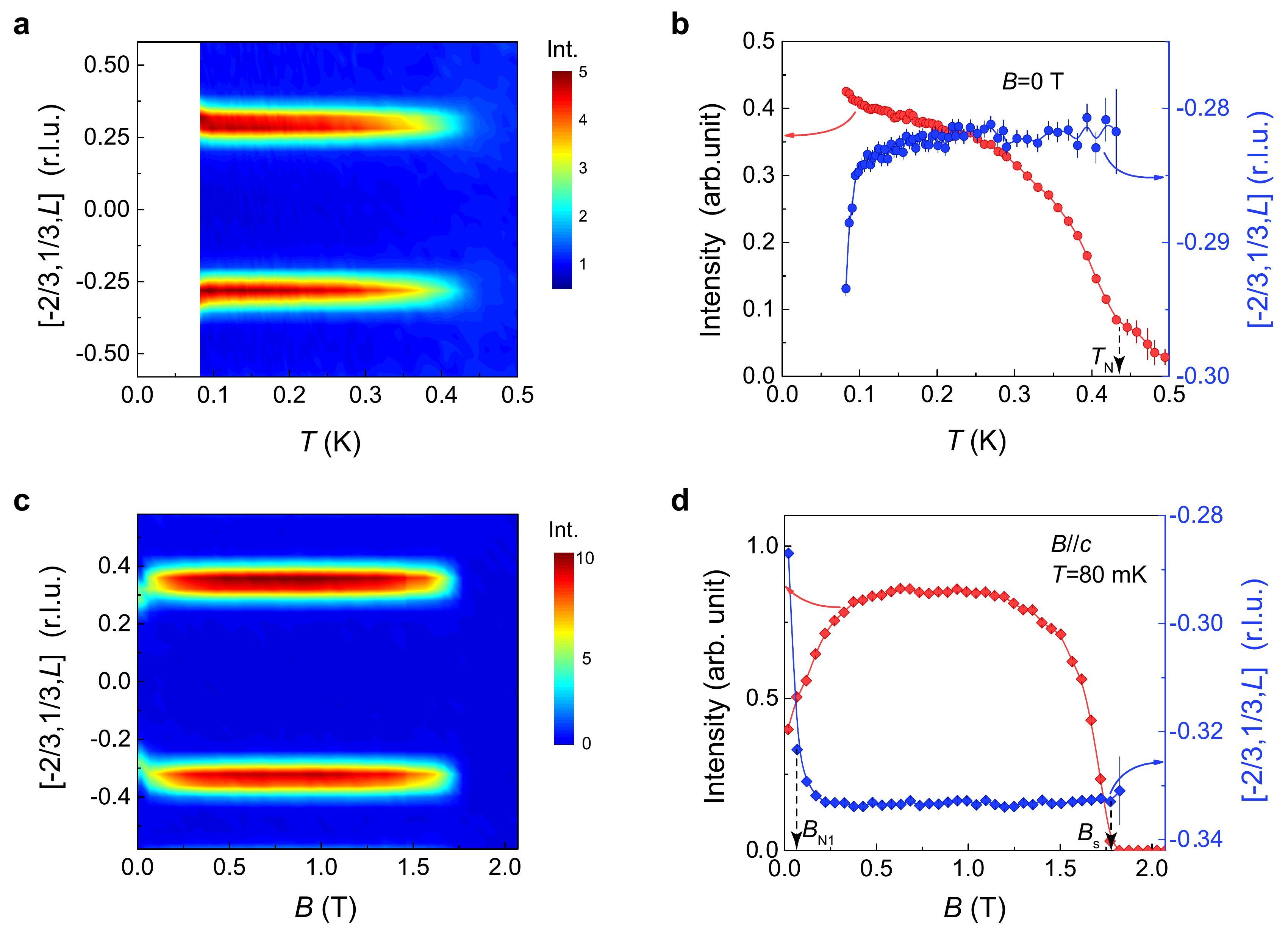}
\caption{{{\bf a} Contour plot of temperature evolution of the magnetic reflections at $(-\frac{2}{3},\frac{1}{3},L)$ measured at zero field. {\bf b} Line cuts of temperature evolution of the incommensurate magnetic reflections ($-\frac{2}{3}$,$\frac{1}{3}$,$L$) measured at zero field. The red and blue filled circles represent the integrated intensity and the $L$ component of the magnetic reflection, respectively. {\bf c} Contour plot of field evolution of the magnetic reflections at $(-\frac{2}{3},\frac{1}{3},L)$ measured at $T=\qty{80}{mK}$, where the magnetic field swept up continuously at a rate of \qty{0.01}{T}/min. {\bf d} Line cuts of field evolution of the incommensurate magnetic reflections $(-\frac{2}{3},\frac{1}{3},L)$ measured at $T=\qty{80}{mK}$. The red and blue diamonds represent the integrated intensity and the $L$ component of the magnetic reflection, respectively. Error bars indicate standard deviations in {\bf b} and {\bf d}.}}
  \label{Order_para}
\end{figure}

Figures~\ref{Order_para}{\bf a}-{\bf b} show the temperature evolution of magnetic reflections ($-\frac{2}{3}$,$\frac{1}{3}$,$L$) at zero field. As the temperature rises from \qty{80}{mK}, the absolute value of the $L$ component dramatically decreases (solid blue dots), indicating that it is near the border of a phase transition. In this zero-field case, we did not reach a lower temperature to fully pin down the transition (similar to the other thermodynamic measurements), indicating that $T_{\rm N1}$ is slightly below \qty{80}{mK}.
Upon further increasing the temperature, the $L$ location of the magnetic reflections remains basically unchanged, whereas the intensity of the magnetic peaks gradually decreases. The dashed black arrow around \qty{431}{mK}, shown in Fig.~\ref{Order_para}{\bf b}, corresponds to the AFM phase transition at $T_{\rm N}$. For $T> T_{\rm N}$, we can still observe the weak diffuse scattering along the $\bm{c}$ axis, resulting in the residual intensity as seen in Fig.~\ref{Order_para}{\bf b}.

\begin{figure}[tbp!]
\centering\includegraphics[width=0.55\columnwidth]{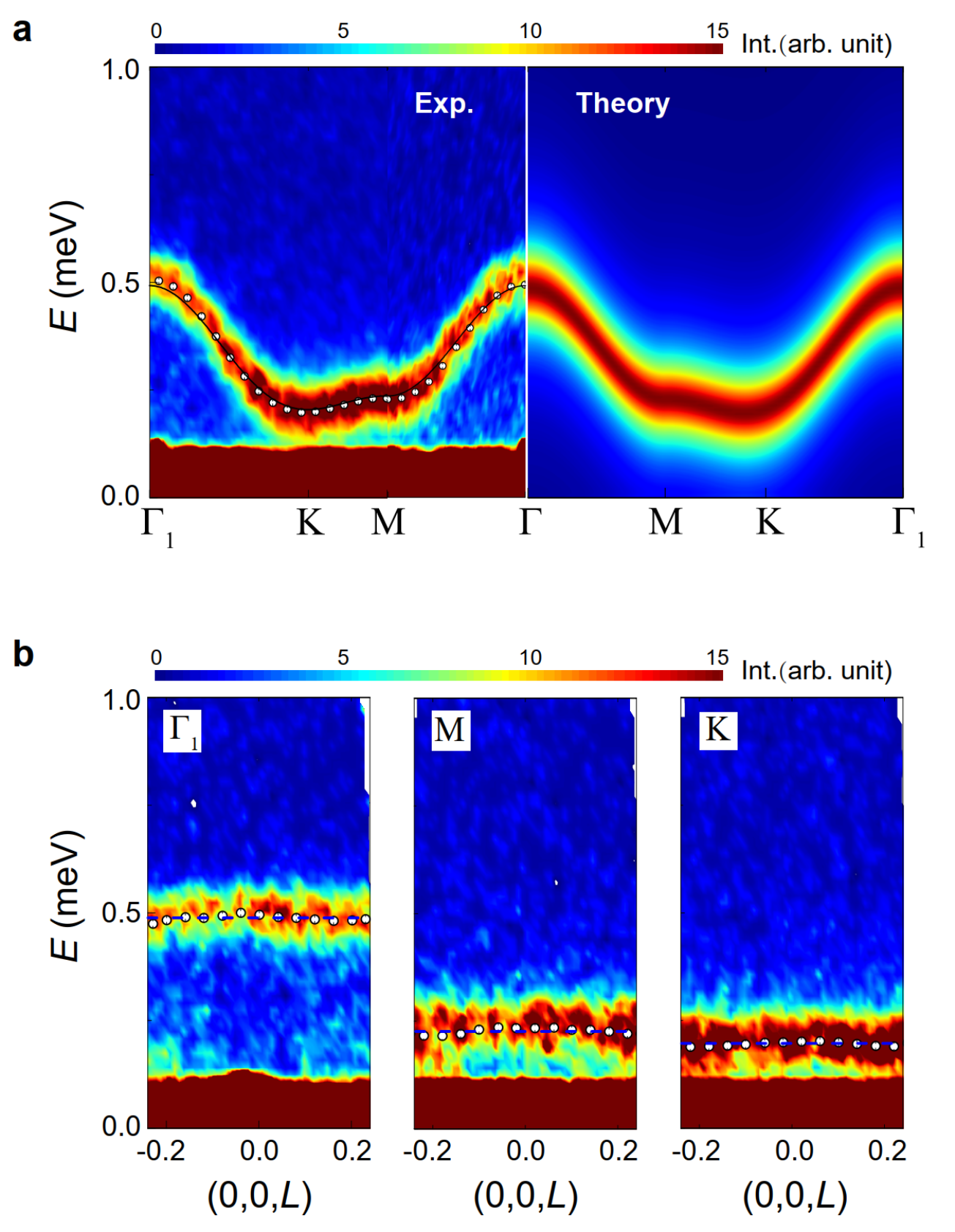} 
\caption{{\bf a} Inelastic neutron scattering intensity measured at $T=\qty{60}{mK}$ and $B=\qty{3}{T}$ along different high symmetry directions. The intensity is integrated over the window of $L=[-0.22, 0.22]$. The black empty circles are centers of the Gaussian fit to the data. The theoretical 1-magnon dispersion (black solid line on the left) and intensity (contour map on the right) used the same parameters as in Fig.~3{\bf c} in the main text, except that here we use $B=\qty{3}{T}$ instead of $B=\qty{5}{T}$. {\bf b} Inelastic neutron scattering intensity along $L$-direction with different in-plane momenta at $T=\qty{60}{mK}$ and $B=\qty{3}{T}$. The black empty circles are centers of the Gaussian fit to the data and the blue dashed lines are guides to the eye. Error bars indicate standard deviations in {\bf a} and {\bf b}.}
\label{EQ_B3T}
\end{figure}

\begin{figure}[tbp!]
\centering
\includegraphics[width=0.95\columnwidth]{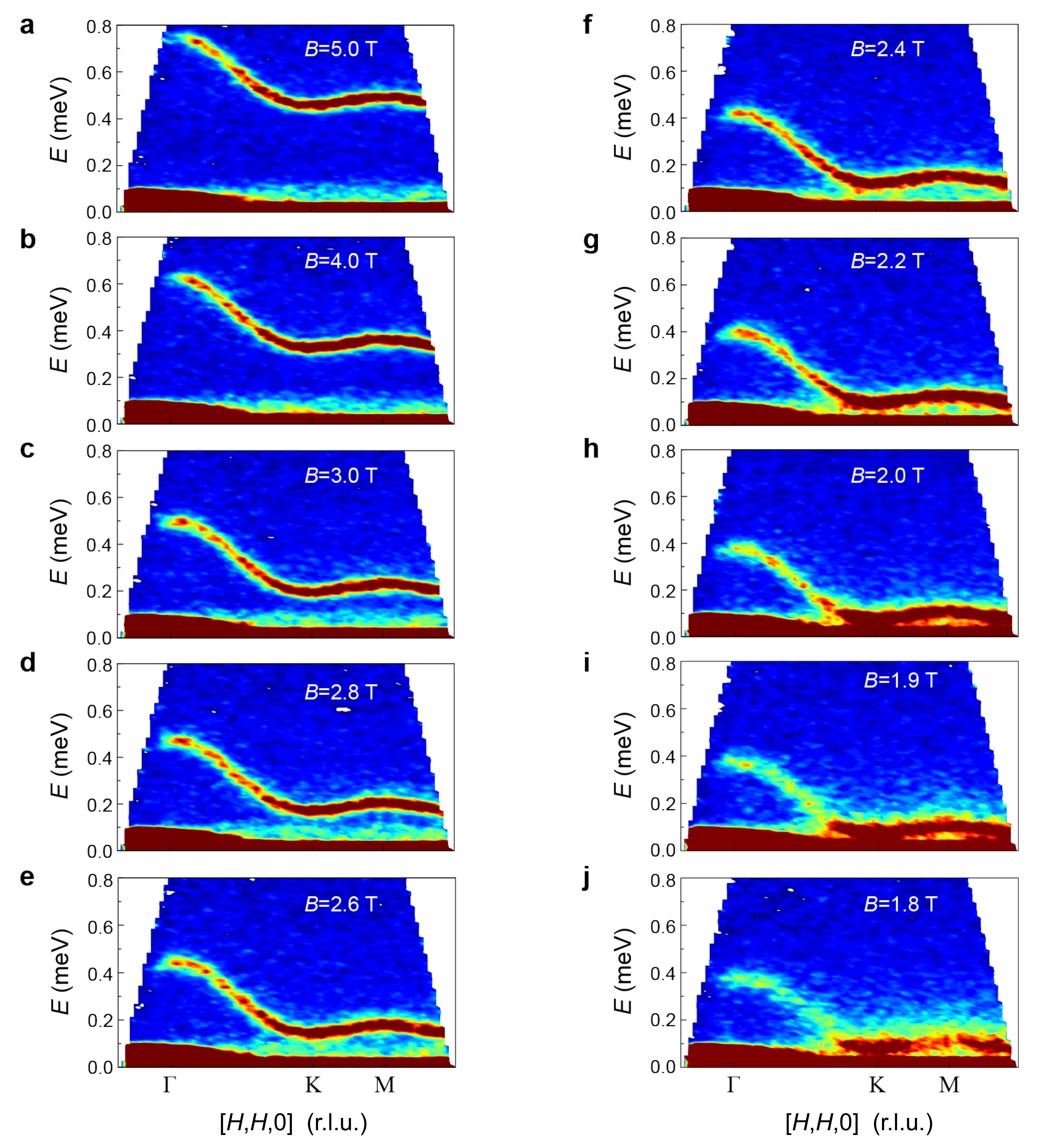}
\caption{{Inelastic neutron scattering spectra along high-symmetry directions measured at $T=\qty{60}{mK}$ and different magnetic fields. The intensity is integrated over the window of $L=[-0.22, 0.22]$. The incident neutron energy is $E_{\rm i}=\qty{1.5}{meV}$.}}
\label{SW}
\end{figure}

To reveal the location of $T_{\rm N1}$, we followed the same strategy as the other thermodynamic measurements performed in this work. By applying a magnetic field along the $\bm{c}$ axis, we indeed see that
the $L$ component of the magnetic reflections undergoes an abrupt change to 1/3 (Fig.~\ref{Order_para}{\bf c}-{\bf d}), confirming that $T_{N1}$ is an incommensurate-commensurate transition, and for $T_{N1}=\qty{80}{mK}$ the corresponding critical field $B_{N1}$ is tiny (see the left arrow in Fig.~\ref{Order_para}{\bf d}).
As the field is further raised, the intensity of the magnetic peak reaches a maximum value in the center of the 1/3-plateau until it is suppressed at the critical field $B_{\rm s}\approx \qty{1.8}{T}$. 

We note that the transition $B_{\rm c2} \approx \qty{1.56}{T}$ was not observed in the neutron diffraction data, where a zero intensity is expected at $(-\frac{2}{3},\frac{1}{3},L)$ in the region $B_\text{c2}<B<B_\text{s}$ where the spin nematic (SN) phase is located.
On the other hand, both the magnetization and MCE data clearly indicate the existence of $B_\text{c2}$.
These seemingly controversial findings are possibly related to the first-order nature of the transition at $B_{\rm c2}$, and the fast continuous field-sweep mode used in the neutron diffraction experiment. 
In fact, even the $T=\qty{350}{mK}$ data were found to exhibit hysteresis for field sweeping up and down, confirming that this discrepancy is an artifact due to the continuous field-sweep mode.

Figure~\ref{EQ_B3T}{\bf a} shows the inelastic neutron scattering spectrum along different high-symmetry directions measured at $T=\qty{60}{mK}$ and $B=\qty{3}{T}$ with the field applied along the $\bm{c}$-axis. Using the same parameters as in Fig.~3{\bf c} in the main text, the theoretical 1-magnon dispersion (black solid line on the left) and intensity (contour map on the right) well reproduced the observed spin wave.

Figure~\ref{SW} shows the inelastic neutron scattering spectra measured at different magnetic fields in the fully polarized (FP) state at $T=\qty{60}{mK}$ using $E_{\rm i}=\qty{1.5}{meV}$. By decreasing the magnetic field, the 1-magnon spin wave gradually moved to lower energies.
An important observation is that the 1-magnon gap never closes across the critical field $B_\text{s}=\qty{1.8}{T}$. 
We caution that the measurement was performed at a finite temperature $T=\qty{60}{mK}$, and the data point at $\qty{1.8}{T}$ is already located inside the quantum critical fan ($T>T^*$), in which region the 1-magnon gap no longer scales linearly as a function of the magnetic field. This behavior is expected regardless of the $U(1)$ symmetry of the model, since inside the critical fan the spin correlation becomes short-ranged and only paramagnons survive.

\newpage

\section{Magnetocaloric Effect (MCE)}

\begin{figure}[bp!]
\centering
\includegraphics[width=0.9\columnwidth]{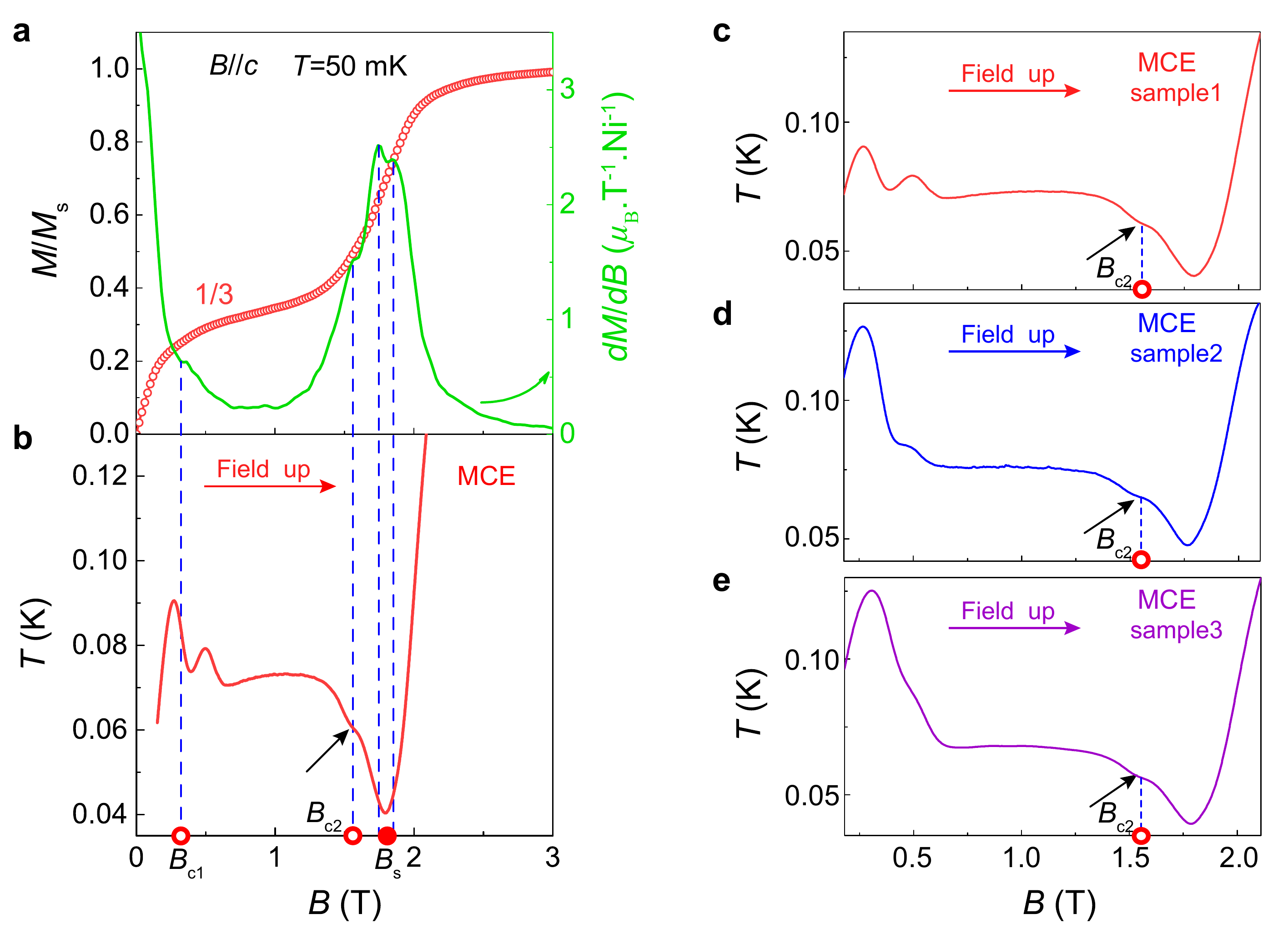}
\caption{{\bf a} Magnetization and magnetic susceptibility $dM/dB$ measured at \qty{50}{mK} with the field applied along the $\bm{c}$ axis. {\bf b} MCE measured by monitoring $T$ while sweeping magnetic field up with a fixed bath temperature $T=\qty{50}{mK}$ and the field applied along the $\bm{c}$ axis. {\bf c}-{\bf e} MCE measured at three different samples with a fixed bath temperature $T=\qty{50}{mK}$ and magnetic field along the $\bm{c}$ axis.}
\label{MCE}
\end{figure}

\begin{figure}[tbp!]
\centering
\includegraphics[width=0.9\columnwidth]{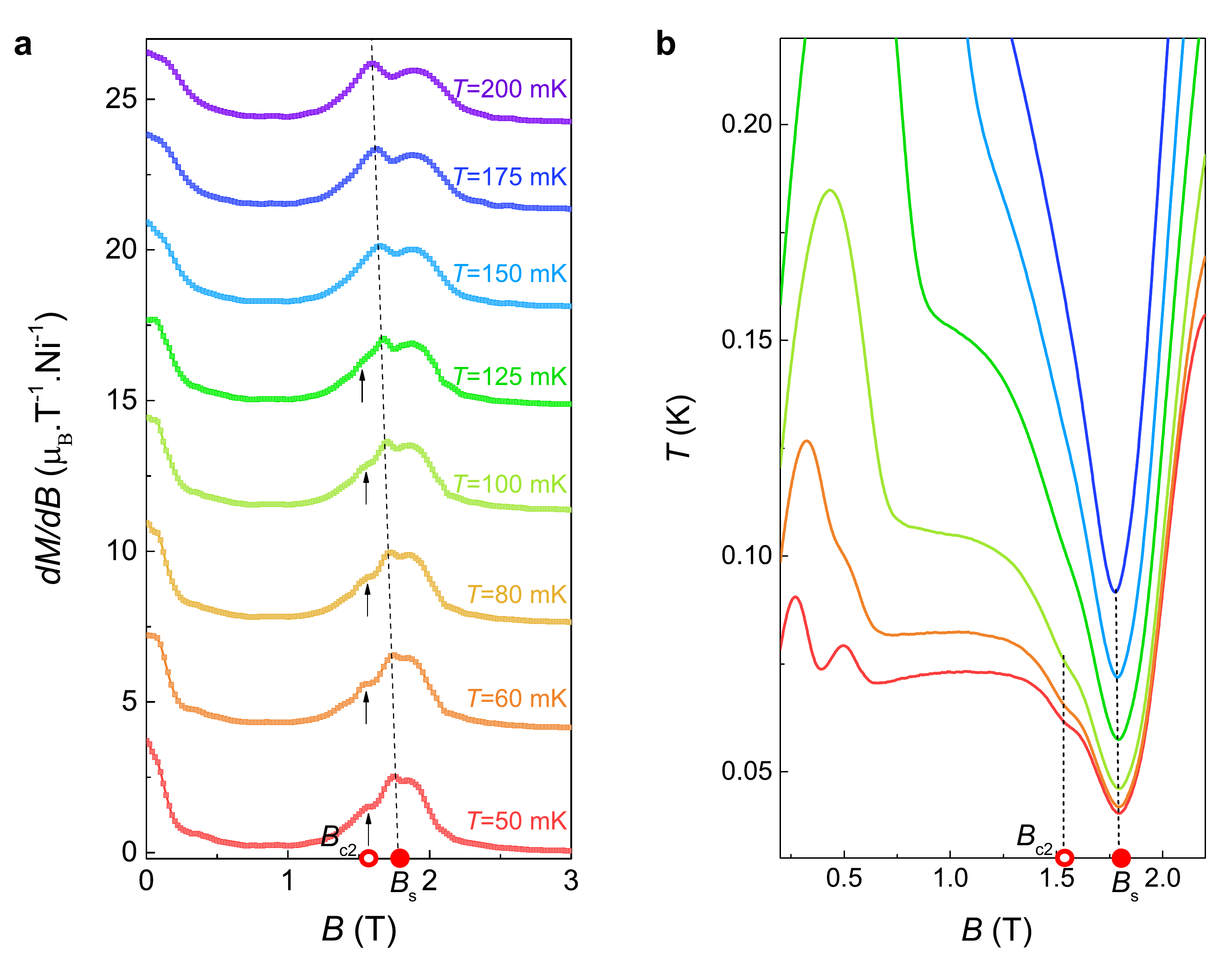}
\caption{{\bf a} Magnetic susceptibility $dM/dB$ measured at different temperatures with magnetic field along the $\bm{c}$ axis. {\bf b} MCE measured by fixing different bath temperatures with magnetic field along the $\bm{c}$ axis.}
\label{MCE_difT}
\end{figure}

As shown in Fig.~\ref{MCE}, the MCE measurements show additional evidence of the transition at $B_{\rm c2}$. By sweeping the magnetic field with a fixed bath temperature $T=\qty{50}{mK}$, the sample temperature variation was recorded as the field was swept through the phase boundaries. It is clear that the value of $B_{\rm c2}$ determined from $dM/dB$ agrees well with the anomaly observed in the MCE curves (Fig.~\ref{MCE}{\bf a}-{\bf b}). To further verify the transition at $B_{\rm c2}$, the MCE measurements were repeated in three different samples. As illustrated in Fig.~\ref{MCE}{\bf c}-{\bf e}, the anomaly at $B_{\rm c2}$ can be observed in all three samples.

Furthermore, the $B_{\rm c2}$ anomaly shows a clear temperature dependence in both magnetization and MCE measurements. As shown in Fig.~\ref{MCE_difT}, we find that the $B_{c2}$ anomaly in $dM/dB$ and MCE curves gradually disappears as we raise the temperature.

\newpage
~
\newpage

\section{ESR spectra}

\begin{figure}[bp!]
\centering
\includegraphics[width=0.95\columnwidth]{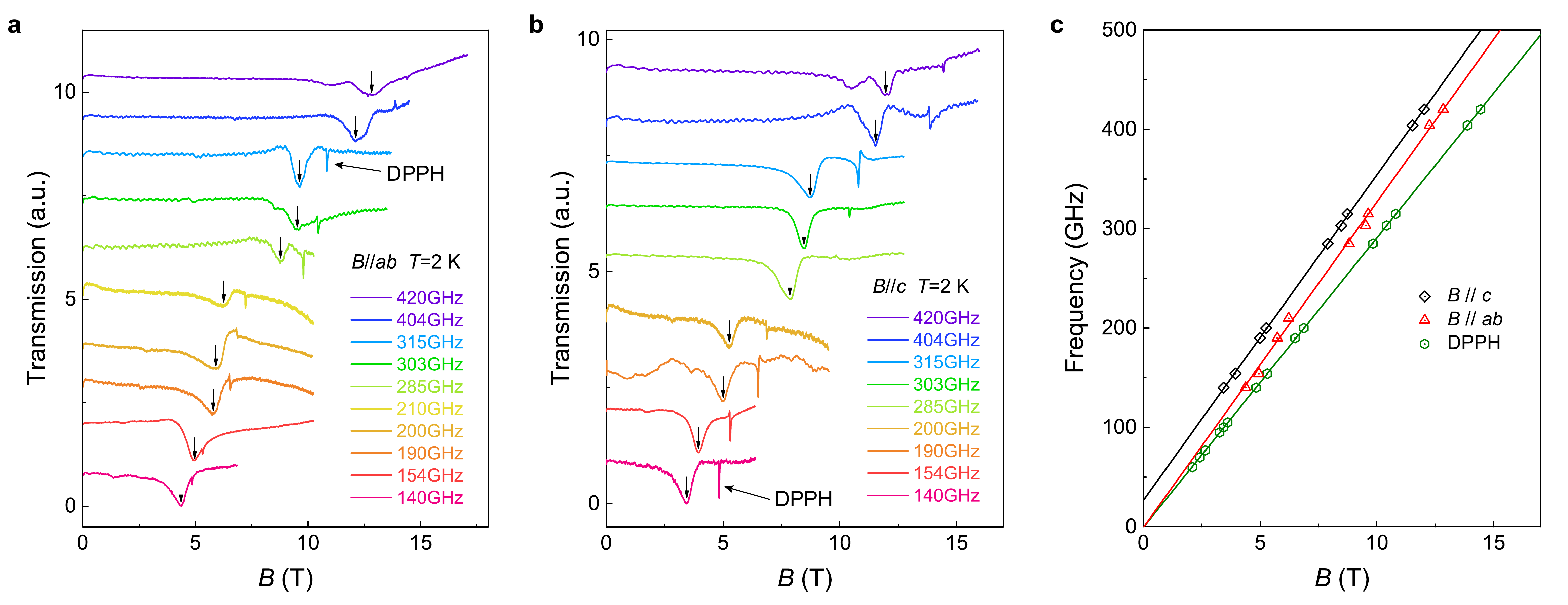}
\caption{{\bf a} and {\bf b} ESR spectra measured at $T=\qty{2}{K}$ in a pulsed magnetic field for $B \parallel \bm{ab}$ and $B \parallel \bm{c}$, respectively. The signals marked by the black arrows are from 1-magnon excitation at the $\Gamma$ point, and the sharp anomalies in the ESR spectra are the resonant signals for DPPH. {\bf c} Frequency-field relation obtained at $T=\qty{2}{K}$ with $B \parallel \bm{ab}$ (red triangles) and $B \parallel \bm{c}$ (black diamonds). The green hexagons are the ESR signals of DPPH.}
\label{ESR_pulsed}
\end{figure}

\begin{figure}[tbp!]
\centering
\includegraphics[width=0.95\columnwidth]{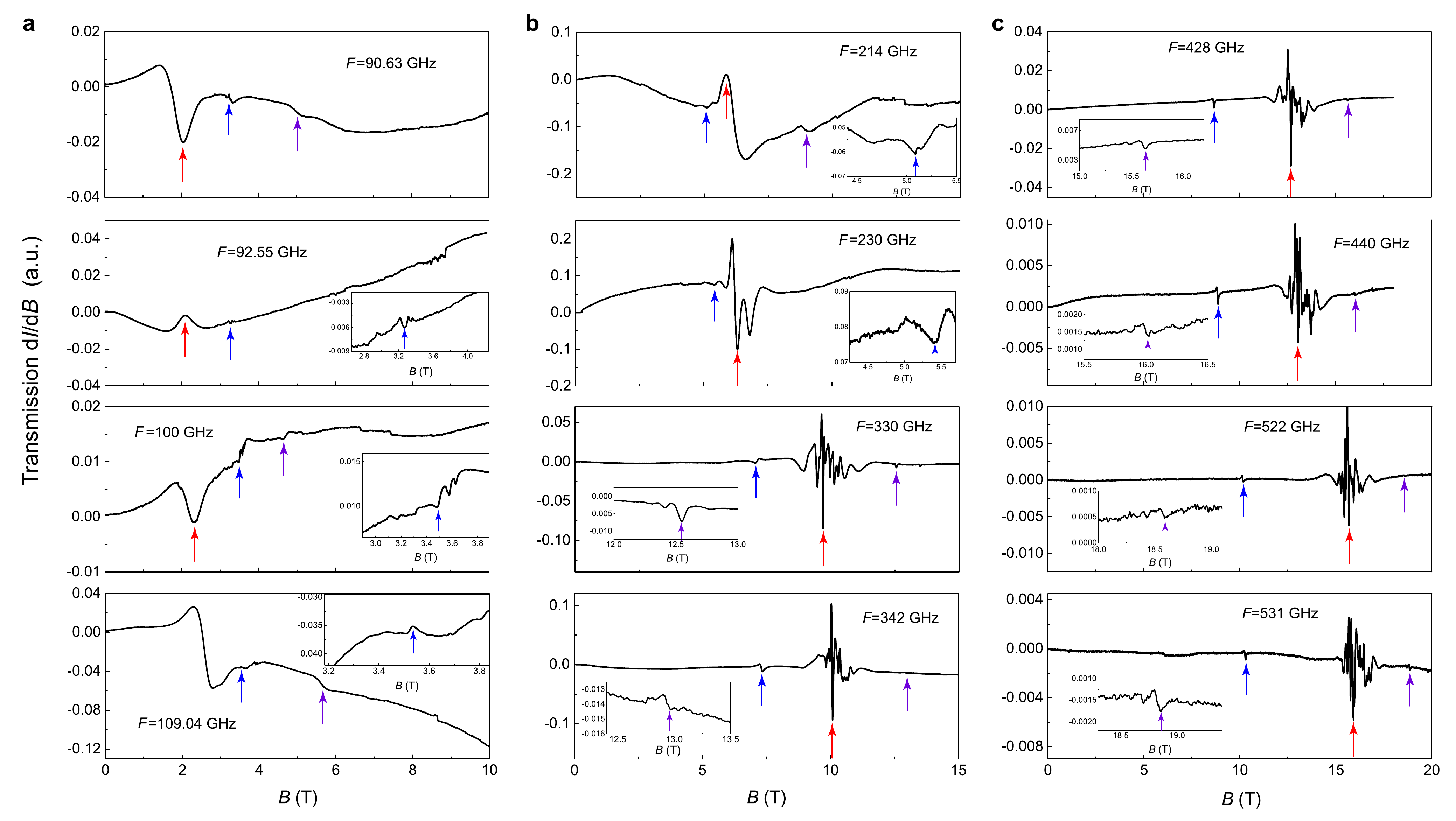}
\caption{ESR spectra measured at $T=\qty{2}{K}$ in a steady magnetic field tilted $6^\circ$ away from the $\bm{c}$ axis. The signals marked by the red, blue and purple arrows represent the 1-magnon excitation at the $\Gamma$ point, 2-magnon bound state at the $\Gamma$ point, and the finite-$T$ excitation from the 1-magnon state to the 2-magnon bound state at the K point, respectively. The insets show zoomed-in view of the excitations.}
\label{ESR_steady}
\end{figure}

\begin{figure}[tbp!]
\centering
\includegraphics[width=0.9\columnwidth]{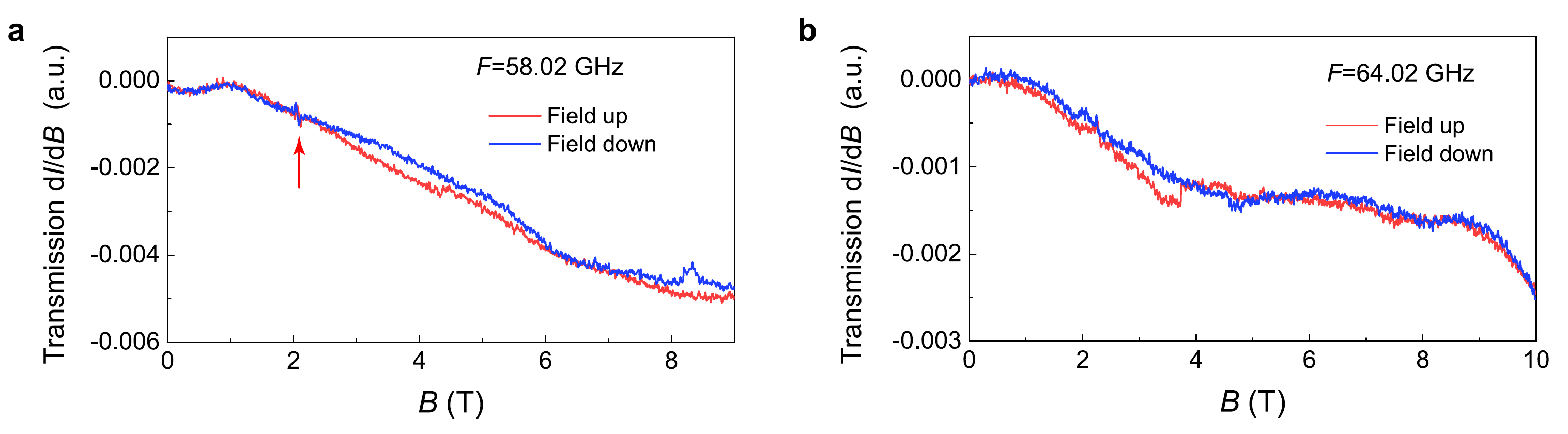}
\caption{ESR spectra measured at $T=\qty{2}{K}$ with frequencies {\bf a} $F=\qty{58.02}{GHz}$; {\bf b} $F=\qty{64.02}{GHz}$ in a steady magnetic field tilted $6^\circ$ away from the $\bm{c}$ axis. The signal marked by the red arrow indicates the 1-magnon excitation at the $\Gamma$ point.}
\label{ESR_F58-64Ghz}
\end{figure}

\begin{figure}[tbp!]
\centering
\includegraphics[width=0.95\columnwidth]{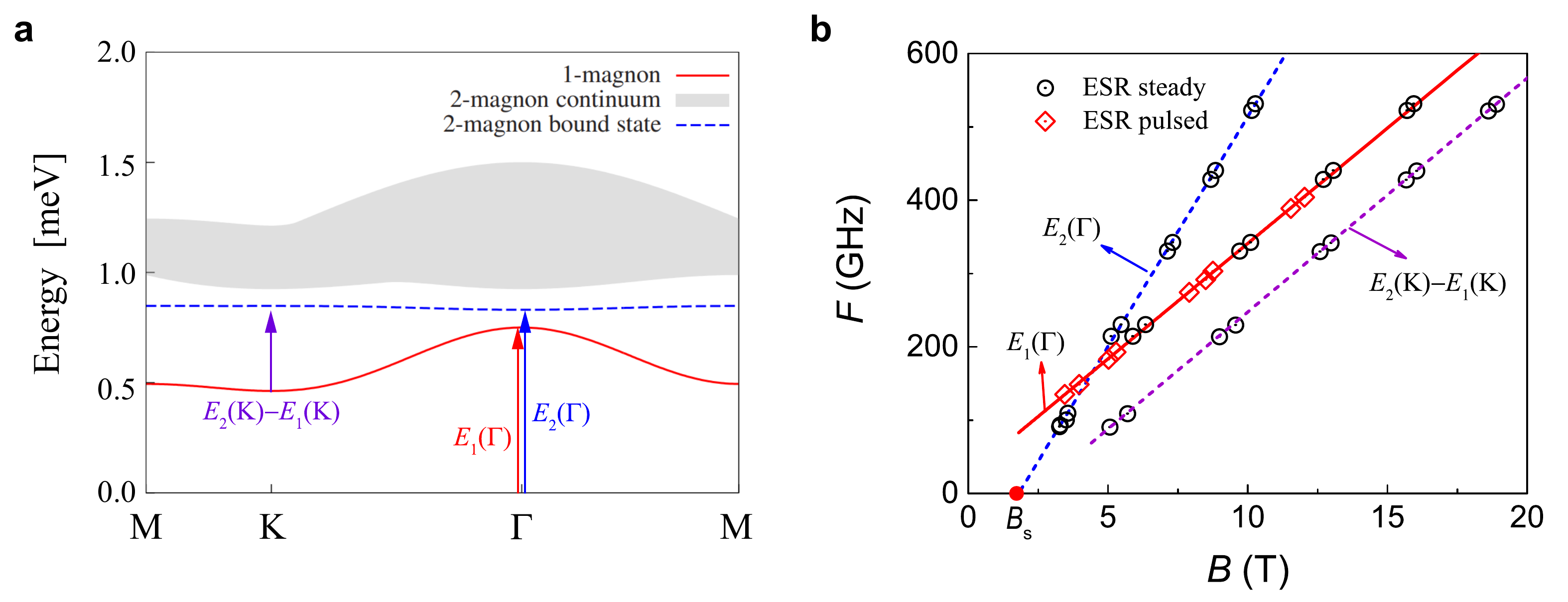}
\caption{{\bf a} Energies of the 1-magnon excitation, 2-magnon continuum and 2-magnon bound state at $B=\qty{5}{T}$, for the TL model (4) in the main text with $J=\qty{0.032}{meV}$, $\Delta=1.13$, $D/J=3.97$, and $g_{c}=2.24$. The  arrows illustrate the dominant signals observed in our ESR experiments. {\bf b} The calculated frequency-field relation of the 1-magnon state (red solid line) and the 2-magnon bound state (blue dashed line) at the $\Gamma$ point, and the finite-$T$ excitation from the 1-magnon state to the 2-magnon bound state at the K point (purple dashed line), with $B \parallel \bm{c}$. The red diamonds and black circles are the observed ESR signals with $B\parallel \bm{c}$ (pulsed) and $B$ tilted $6^\circ$ away from the $\bm{c}$-axis (steady), respectively.}
\label{ESR_F_dep_B}
\end{figure}

As shown in Figs.~\ref{ESR_pulsed}{\bf a}-{\bf b}, our earlier ESR measurements were carried out in a pulsed magnetic field at $T=2$ K with $B \parallel \bm{ab}$ and $B \parallel \bm{c}$, respectively. Resonance peaks from 1-magnon excitation at $\Gamma$ point marked by the black arrows are observed at different frequencies for both field directions. The corresponding frequency-field ($F-B$) relations for $B \parallel \bm{c}$ and $B \parallel \bm{ab}$ are shown in Fig.~\ref{ESR_pulsed}{\bf c}. By linear fittings, we can get the $g$-factors $g_c=2.25(1)$ and $g_{ab}=2.24(5)$ (corrected by the measured $g_{\rm {DPPH}}$), respectively. 

Due to the noisy signals in our pulsed-field ESR spectra, we did not observe the weak 2-magnon bound state there. Later, we performed another ESR experiment in a steady magnetic field, where the magnetic field was tilted $6^\circ$ away from the $\bm{c}$ axis. With the first-derivative collecting mode, higher quality data revealed more details in the steady-field ESR spectra. Figure~\ref{ESR_steady} shows the ESR spectra of \NiP{} measured in a steady magnetic field at $T=\qty{2}{K}$. Besides the 1-magnon excitations at $\Gamma$ point (indicated by the red arrows in Fig.~\ref{ESR_steady}), we can also observe the weak signals of the 2-magnon bound state at $\Gamma$ point (blue arrows) and the finite-$T$ excitations from the 1-magnon state to the 2-magnon bound state at the $K$ point (purple arrows at higher field) between 90.63 GHz and 531 GHz. 
The specific excitation paths are also illustrated in Fig.~\ref{ESR_F_dep_B}{\bf a}.  By linear fitting of the ESR data, the critical field of the BEC of 2-magnon bound state was determined as $\qty{1.80(3)}{T}$ (shown in Fig.~\ref{ESR_F_dep_B}{\bf b}), which exactly corresponds to the observed two-dimensional QCP around $B_\text{s}=\qty{1.80}{T}$ based on our magnetization and specific heat measurements. We note that there are also a series of resonance signals around the 1-magnon resonance peaks, possibly corresponding to the finite-$T$ excitations from the 1-magnon state to the 2-magnon continuum.

At lower frequencies, as shown in Fig.~\ref{ESR_F58-64Ghz}, we observed a significant decrease in the signals of magnon excitations. In particular, the 2-magnon bound state becomes indiscernible as it is overwhelmed by the background noise. This is mainly a finite-temperature effect, given that our ESR measurements were performed at \qty{2}{K}. 
In fact, the data point with a frequency of \qty{90.63}{GHz} is already located at $B=\qty{3.24}{T}$, which is slightly inside the critical fan region ($T>T^*$).
Clearly, at this temperature any 2-magnon bound state signal at lower frequency becomes physically ill-defined.

The lines in Fig.~\ref{ESR_F_dep_B}{\bf b} are the calculated $F-B$ relation of the 1-magnon state (red solid line) and the 2-magnon bound state (blue dashed line) at the $\Gamma$ point, and the finite-$T$ excitation from the 1-magnon state to the 2-magnon bound state at the K point (purple dashed line), with $B \parallel \bm{c}$. The red diamonds and the black circles are the ESR signals with $B \parallel \bm{c}$ (pulsed) and $B$ tilted $6^\circ$ away from the $\bm{c}$ axis (steady), respectively. 

\newpage

\section{NMR measurements}

\begin{figure}[hbp!]
\centering
\includegraphics[width=0.95\columnwidth]{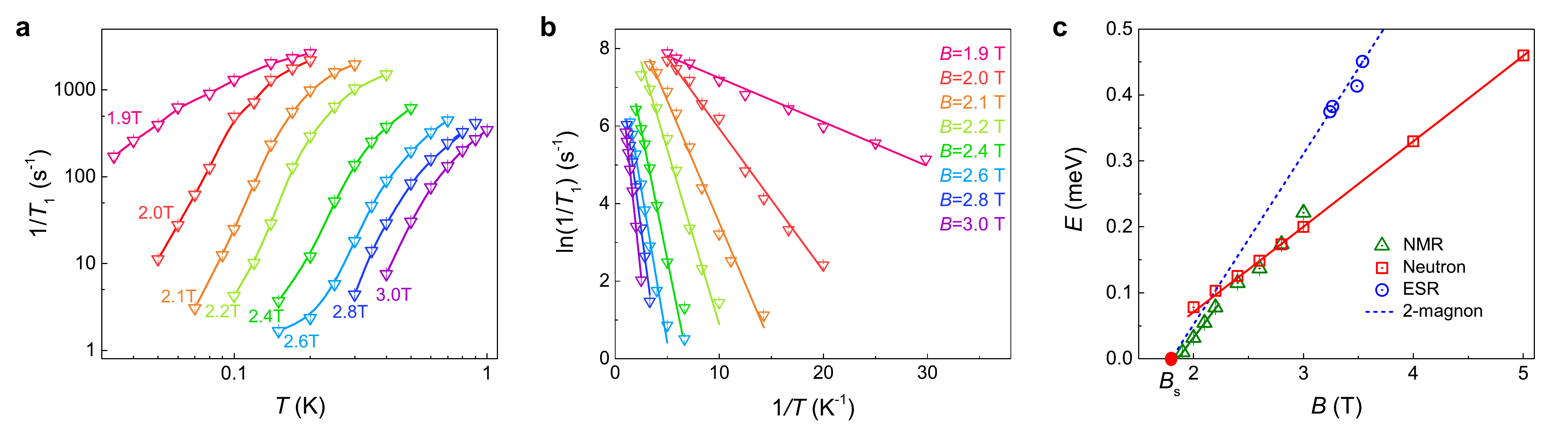}
\caption{{\bf a} Spin-lattice relaxation rate $1/T_1$ as a function of the temperature measured at different fixed fields with $B \parallel \bm{c}$.
The solid lines are guides to the eye.
{\bf b} Same data plotted in logarithmic scale. The solid lines are the linear fitting to the data with $\ln(1/T_1) \sim -\dfrac{\Delta}{T}$. {\bf c} The calculated energy-field relation of the 1-magnon state (red solid line) at the K point and the 2-magnon bound state (blue dashed line) at the $\Gamma$ point with $B \parallel \bm{c}$. The red squares and blue circles are the observed 1-magnon state by neutron scattering and 2-magnon bound state by ESR, respectively. The green triangles are the energy gap according to NMR measurements. Error bars indicate standard deviations.}
\label{NMR_gap}
\end{figure}

The $^{23}$Na NMR measurements were performed at temperatures down to \qty{30}{mK} to locate the energy of the magnon gap near the critical field. Figure~\ref{NMR_gap}{\bf a} shows the temperature dependence of the spin-lattice relaxation rate measured at different magnetic fields above $B_\text{s}$. 
A fit to the function $1/T_1 \propto \exp (- \Delta /T)$ (see Fig.~\ref{NMR_gap}{\bf b})
reveals the evolution of the magnon gap as a function of the magnetic field (see Fig.~\ref{NMR_gap}{\bf c}).

\newpage

\section{3- and 4-magnon states}
To exclude the possibility of a first-order transition at $B_{\rm s}$, we check whether the 2-magnon bound states have attractive interactions, i.e., whether there is formation of 3- or 4-magnon bound states in the relevant parameter region. This can in principle be calculated exactly by solving the Lippmann-Schwinger equation, similarly to the 2-magnon solution presented in the main text. However, this approach quickly becomes over complicated as we increase the magnon number. In this section, we use a different method that can be easily generalized to handle larger numbers of magnons. The method was initially proposed by Trugman to study the holes in high-temperature superconductors~\cite{TrugmanSA1988,TrugmanSA1990}, and was later generalized to other systems~\cite{ElShawishS2006,HaravifardS2006,WangZ2018a}.

\begin{figure*}[tbp!]
\includegraphics[width=1.0\textwidth]{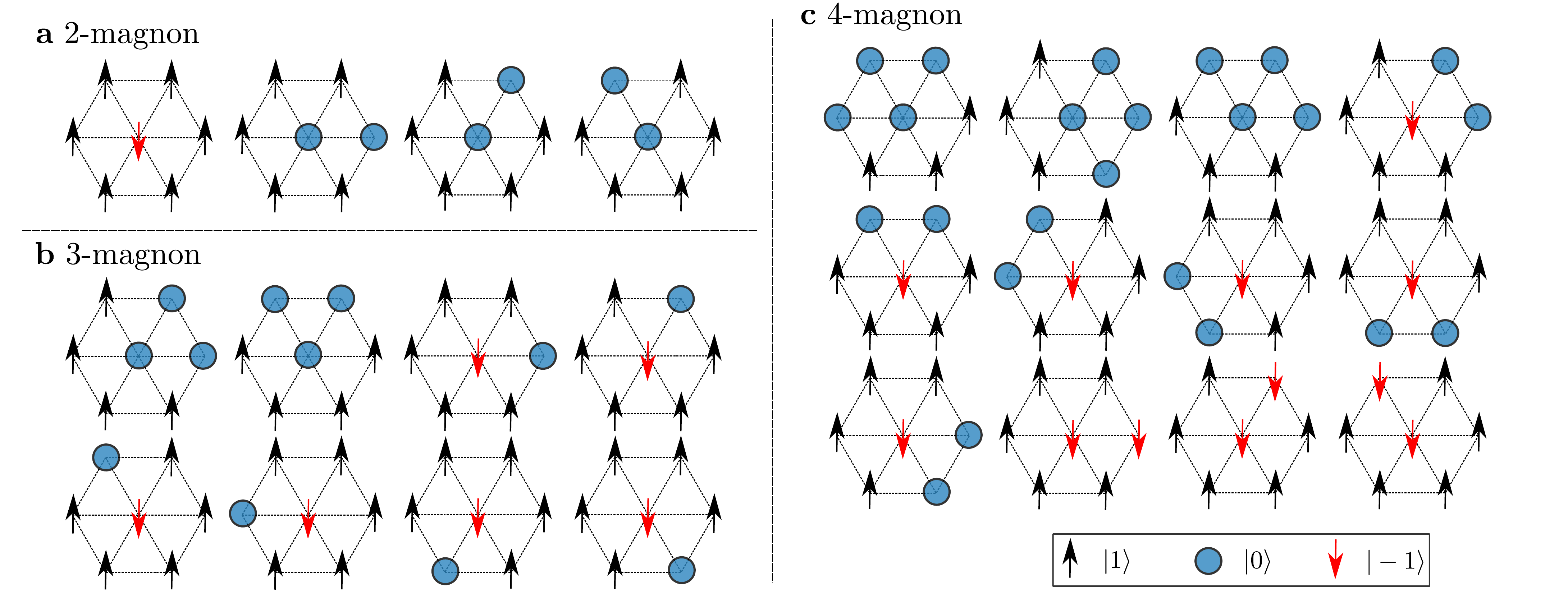}
\caption{{\bf Initial basis for 2-, 3-, and 4-magnon states used in the variational method.}}
\label{fig:initial}
\end{figure*}

Here we outline the basic idea of the method.
In the FP state, the excitations are gapped for $B>B_{\rm s}$. As a consequence, the size of the excited states is controlled by the finite correlation length $
\xi$.
Motivated by this observation, we start by creating local excitations $|\varphi_i\rangle$ (see Fig.~\ref{fig:initial} for the choice of initial states) and project them to momentum space:
\begin{equation}
|\varphi_i (\bm{k})\rangle \equiv \frac{\hat{P}(\bm{k}) |\varphi_i \rangle}{\sqrt{\langle \varphi_i | \hat{P}(\bm{k}) |\varphi_i \rangle}}.
\end{equation}
Here $\hat{P}(\bm{k}) \equiv \frac{1}{N} \sum_{\bm{r}}e^{\iu \bm{k} \cdot \bm{r}} \hat{T}(\bm{r})$ is the projection operator.
Then we apply the Hamiltonian $\mathcal{H}_\text{TL}$ to $|\varphi_i (\bm{k}) \rangle$ to generate new states that dress up $|\varphi_i (\bm{k}) \rangle$. Systematic improvement of the variational space is achieved by repeating this process iteratively. Finally, we diagonalize $\mathcal{H}_\text{TL}$ in the variational basis \{$\varphi_i (\bm{k})$\} to obtain both energy and wave function of the excited states. The convergence of this approach can be checked by comparing the energies and the measured quantities at different iterations.

\begin{figure*}[t!]
\includegraphics[width=1.0\textwidth]{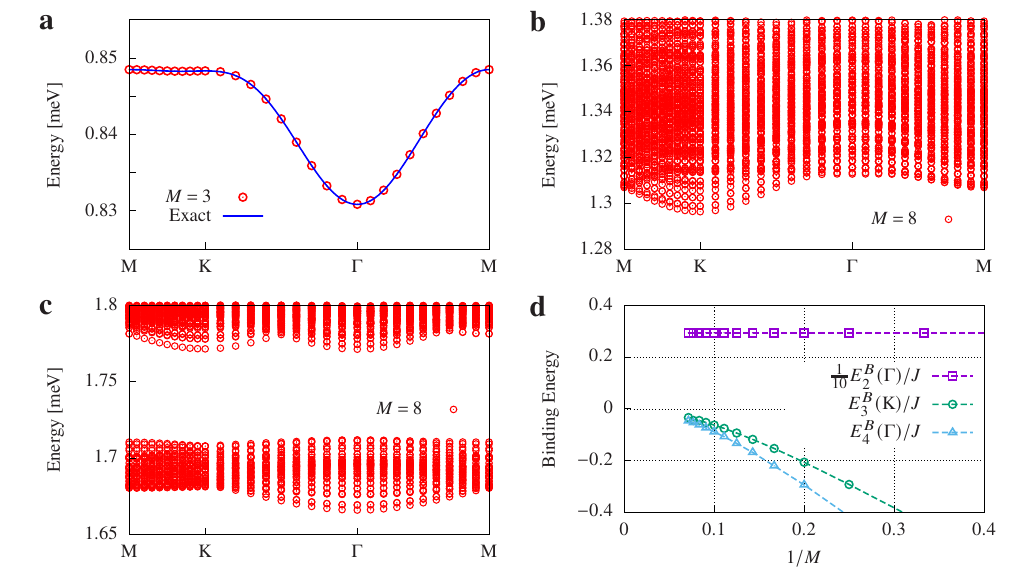}
\caption{{\bf Results of the variational calculation for $\mathcal{H}_\text{TL}$.} The parameters are fixed at $J=\qty{0.032}{meV}$, $\Delta=1.13$, $D/J=3.97$, $g_c=2.24$ and $B=\qty{5}{T}$. {\bf a} Comparison of the variational calculation at $M=3$ and the exact results for the 2-magnon bound state. {\bf b} Dispersion of the 3-magnon states at $M=8$. {\bf c} Dispersion of the 4-magnon states at $M=8$. {\bf d} Binding energies of the 2-, 3-, and 4-magnon states at different iterations. }
\label{fig:234magnon}
\end{figure*}

We start by showing that the variational approach indeed agrees with the exact results for the 2-magnon bound state. After $M=3$ iterations from the total $S^z=N-2$ initial basis (see Fig.~\ref{fig:initial}{\bf a}), the dimension of the variational space expands from $\mathcal{D}=4$ to $\mathcal{D}=31$. Figure~\ref{fig:234magnon}{\bf a} shows that the dispersion of the 2-magnon bound state practically agrees with the exact solution even at $M=3$. In other words, the 2-magnon bound state occupies a relatively small spatial range. Similarly, the dispersion of the 3- and 4-magnon states at $M=8$ iteration is shown in Figs.~\ref{fig:234magnon}{\bf b} and {\bf c}, respectively. 

The minimum of the 3-magnon state is located at the K-point (Fig.~\ref{fig:234magnon}{\bf b}). The binding energy is defined by subtracting a 2-magnon bound state at $\Gamma$-point and a 1-magnon state at K-point:
\begin{equation}
E_3^B(\text{K}) = \left[ E_2(\Gamma) + E_1(\text{K}) \right] - E_3(\text{K}).
\end{equation}
Figure~\ref{fig:234magnon}{\bf d} shows no evidence that supports formation of a 3-magnon bound state up to $M=14$ iterations, and indicates that $E_3^B(\text{K})\rightarrow 0$ for $1/M\rightarrow 0$. In fact, the continuum of states in Fig.~\ref{fig:234magnon}{\bf b} already indicates that these are all scattering states between a 2-magnon bound state and a 1-magnon state.

Similarly, the minimum of the 4-magnon state is located at the $\Gamma$-point (Fig.~\ref{fig:234magnon}{\bf c}). The binding energy is defined by subtracting two 2-magnon bound states both at $\Gamma$:
\begin{equation}
E_4^B(\Gamma) = 2 E_2(\Gamma) - E_4(\Gamma).
\end{equation}
Again, Figure~\ref{fig:234magnon}{\bf d} indicates that $E_4^B(\Gamma)\rightarrow 0$ for $1/M\rightarrow 0$. In other words, there is no formation of a 4-magnon bound state. Note that the narrow continua of states below \qty{1.72}{meV} in Fig.~\ref{fig:234magnon}{\bf c} are the scattering states of a pair of 2-magnon bound states.

\newpage

\section{DMRG calculation}
We use the same TL model as in the main text:
\begin{equation}
\mathcal{H}_\text{TL}=J\sum_{\langle i,j\rangle}\left( S_i^xS_j^x + S_i^yS_j^y +\Delta S_i^zS_j^z\right)
  -D\sum_i\left(S_i^z \right)^2  -\widetilde{H}\sum_i S_i^z,
  \label{hmlt}
\end{equation}
where $S_i^{\alpha}$ ($\alpha=x,y,z$) are the spin-1 operators, and
$\langle i,j\rangle$ runs over the nearest neighbor bonds. 
The anisotropies were fixed as $\Delta=1.13$ and $D/J=3.97$. 
In this section, we set $J=1$ as the unit of energy, and the results here should be scaled correspondingly for comparison to the main text. The effective magnetic field $\widetilde{H}\equiv
g_c\mu_BB$ has already absorbed the $g$-factor $g_c$ and the Bohr magneton $\mu_\text{B}$.

\begin{figure}[bp!]
\centering
\includegraphics[width=.95\textwidth]{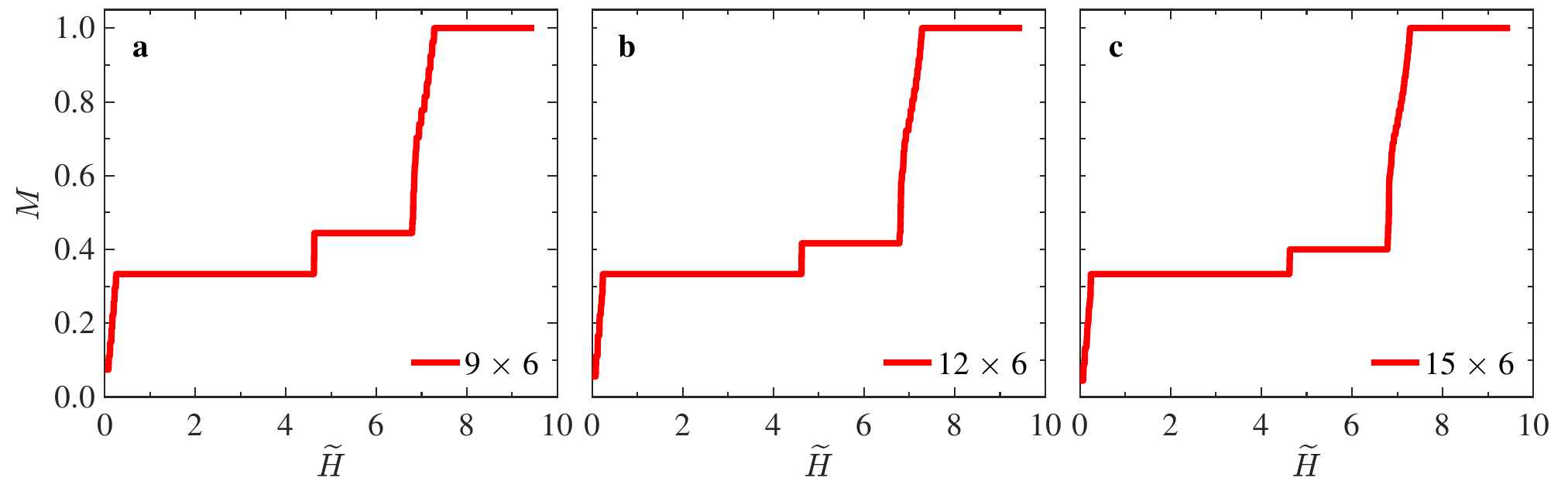}
\caption{Field dependence of the $T=0$ magnetization for cylinders of size {\bf a} $9\times 6$, {\bf b} $12\times 6$, and {\bf c} $15\times 6$.  }
\label{mz1}
\end{figure}

\begin{figure}[bp!]
\centering
\includegraphics[width=.95\textwidth]{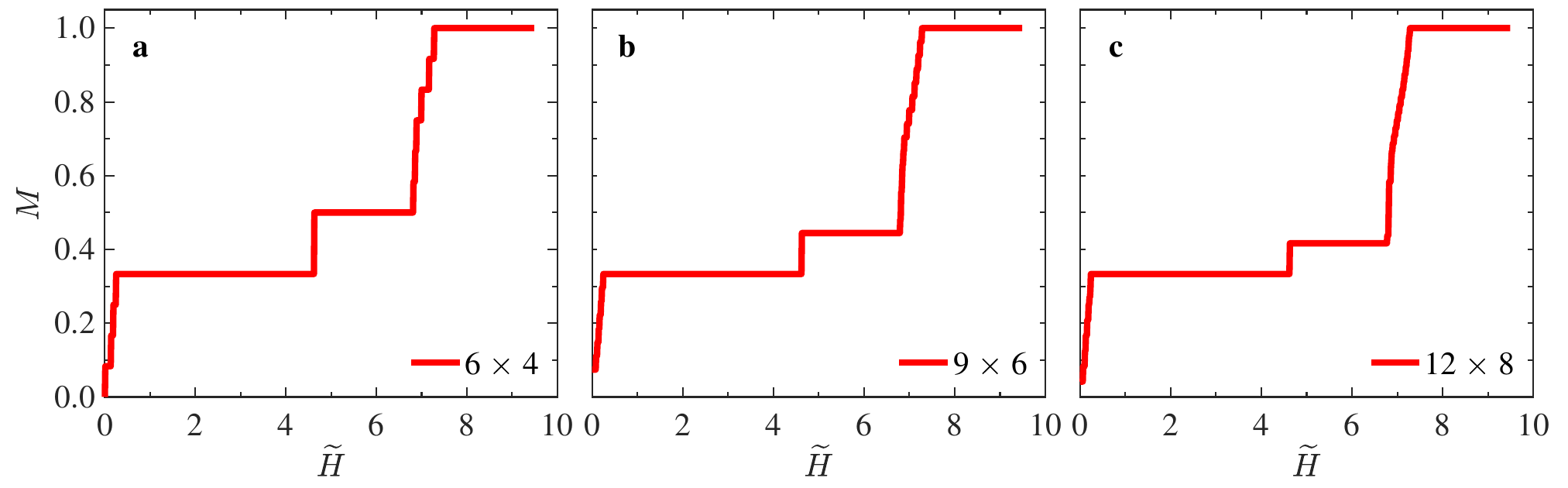}
\caption{Field dependence of the $T=0$ magnetization for cylinders of size {\bf a} $6\times 4$, {\bf b} $9\times 6$, and {\bf c} $12\times 8$.  }
\label{mz2}
\end{figure}

The Hamiltonian \eqref{hmlt} was studied using a U(1)
symmetric density matrix renormalization group (DMRG) method on long
cylinders and torus. 
We first took $\widetilde{H}=0$, and performed the ground state search procedure
for every total spin quantum sector $S_z=\sum_i S_i^z$ on a $L_x\times
L_y$ cylinder ($y$ direction periodic and $x$ direction open) or
torus, with $S_z\in [0,N]$, where $N=L_xL_y$ is the total number of lattice sites. When the external field
$\widetilde{H}>0$, the total energy of each spin sector $S_z$ is
shifted down by $\widetilde{H}S_z$ from its zero-field value,
{\it i.e.},  $E(S_z,\widetilde{H})=E(S_z,0)-\widetilde{H}S_z$. The
$T=0$ magnetization can be
obtained as $M=S_z^m/N$, where $S_z^m$ is the total spin sector
which takes the lowest energy among all sectors at any given
$\widetilde{H}$. In other words,
$E(S_z^m,\widetilde{H})=\text{min}\{E(S_z,\widetilde{H})\}$.

\begin{figure}
\centering
\includegraphics[width=.4\textwidth]{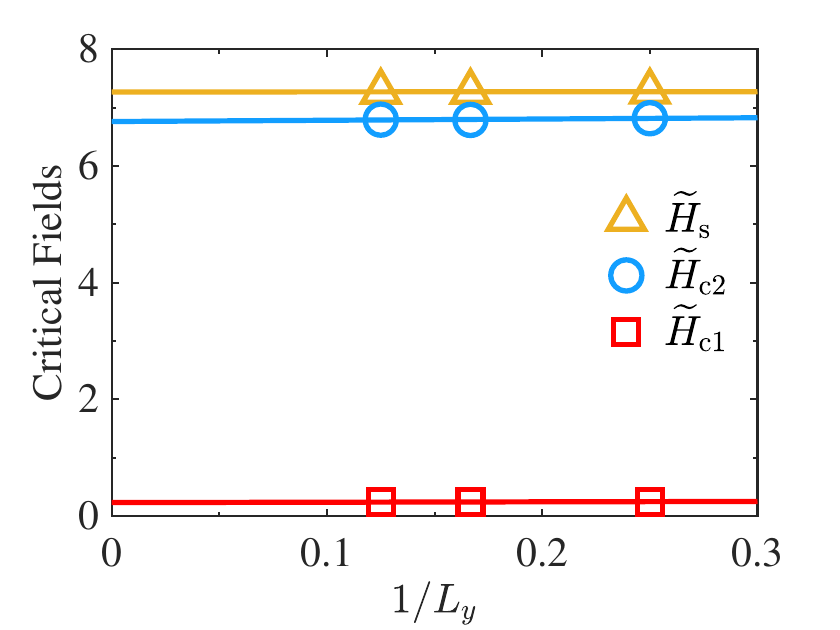}
\caption{Extrapolation of critical fields $\widetilde{H}_{\rm c1}$, $\widetilde{H}_{\rm c2}$, and $\widetilde{H}_{\rm s}$ as a function of $1/L_y$ to the thermodynamic limit, where linear regressions were used.}
\label{extrapolateH}
\end{figure}

In addition to the results on a torus presented in Fig.~5{\bf a} in the main text, we computed the $T=0$ magnetization curve on two series of
cylinders. The first series involves cylinders of fixed width $L_y=6$
but of different length $L_x=9,12,15$. A maximum bond dimension of $1200$
was used to ensure the convergence of the ground state energy in each
quantum sector $S_z$. The ground state energy $E(S_z,0)$ of each sector was
extrapolated to zero discarded weight $\epsilon$, defined as the sum
of the discarded eigenvalues of the reduced density matrix. In the worst
cases, $\epsilon$ reaches $10^{-5}$, and in other cases $\epsilon$ is
around $10^{-8}$. In Fig.~\ref{mz1}, we show magnetization as a
function of $\widetilde{H}$ for the $9\times 6$, $12\times 6$, and
$15\times 6$ cylinders. We find that all three sizes exhibit a wide
$1/3$ plateau starting from $\widetilde{H}_{\rm c1}\approx 0.23$, ending around
$\widetilde{H}=4.8$. This $1/3$-plateau is followed by another
``artificial'' plateau whose magnetization $M$ drops systematically
with increasing $L_x$.
We note that the total spin quantum number difference
of the two neighboring plateaus is a constant ($\Delta S_z=L_y$) for any
given $L_y$, therefore we believe that the ``artificial'' plateau
will eventually merge into the $1/3$ plateau in the thermodynamic
limit, whose range expands from $\widetilde{H}_{\rm c1}\approx 0.23$ to
$\widetilde{H}_{\rm c2}\approx 6.76$. Following a narrow SN phase for
$\widetilde{H}\in [6.76,7.27]$, the magnetization curve finally
saturates when $\widetilde{H}\ge \widetilde{H}_{\rm s}\approx 7.27$.

In the second series of lattices, we increase the length $L_x$ and the
width $L_y$ proportionally, using cylinders of sizes $6\times 4$,
$9\times 6$, and $12\times 8$. We aim to accurately extrapolate the critical
field strengths $\widetilde{H}_{\rm c1}$, $\widetilde{H}_{\rm c2}$,
and $\widetilde{H}_{\rm s}$ to the thermal dynamic
limit. The maximum bond dimensions used are 400 for $6\times 4$, 1200
for $9\times 6$, 3000 for $12\times 8$ with the worst discarded weight
reaches a scale of $10^{-5}$. The magnetization curves of these three
sizes are shown in Fig.~\ref{mz2}. To our surprise, all critical
fields $\widetilde{H}_{\rm c1}$, $\widetilde{H}_{\rm c2}$, and
$\widetilde{H}_{\rm s}$ have little finite-size effects, and they
extrapolate almost as a flat line in $1/L_y$ to the thermodynamic
limit, as shown in Fig.~\ref{extrapolateH}. We conclude that
$\widetilde{H}_{\rm c1}=0.230(8)$, $\widetilde{H}_{\rm c2}=6.76(3)$,
and $\widetilde{H}_{\rm s}=7.270(4)$ in the thermodynamic limit. Here
the error bars are defined by the largest absolute difference of the
critical fields among different sizes. To compare with the
experimental results, we convert $\widetilde{H}_{\rm c1}$,
$\widetilde{H}_{\rm c2}$, and $\widetilde{H}_{\rm s}$ to $B_{\rm
  c1}=\qty{0.057(2)}{T}$, $B_{\rm c2}=\qty{1.668(8)}{T}$ and $B_{\rm s}=\qty{1.794(1)}{T}$. 

\begin{figure}
\centering
\includegraphics[width=\textwidth]{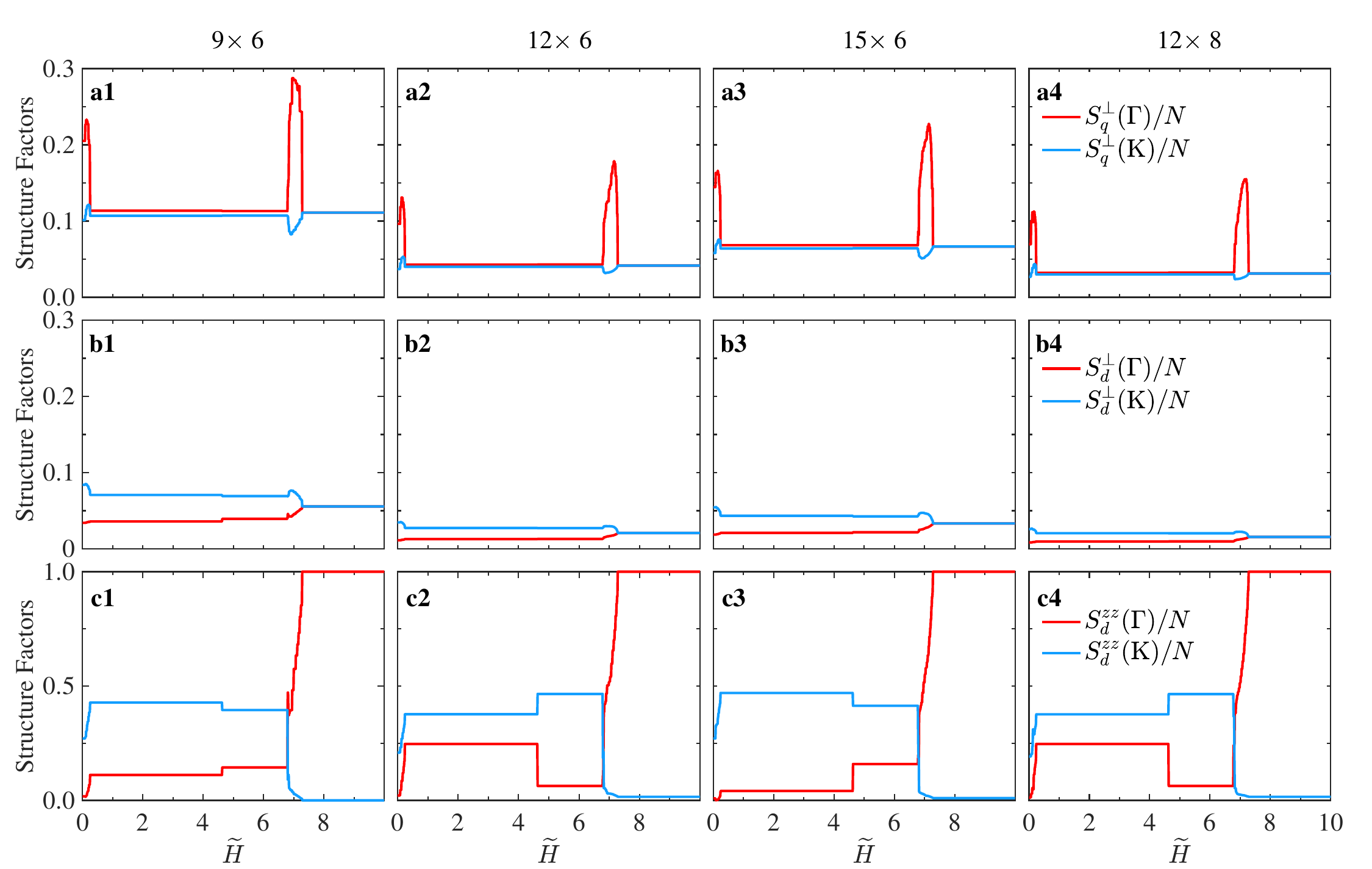}
\caption{{\bf Dipolar and quadrupolar structure factors at $\Gamma$ and K points.} {\bf a} In-plane quadrupolar structure factor, {\bf b} In-plane dipolar structure factor, {\bf c} Out-of-plane dipolar structure factor. The corresponding lattice sizes are indicated on top of each column.}
\label{order1}
\end{figure}

To reveal the nature of these phases, we computed the following structure factors:
\begin{subequations}
\begin{align}
 S^{zz}_d(\bm{k})&=\frac{1}{N}\sum_{i,j}e^{- \iu \bm{k}\cdot(\bm{r}_i-\bm{r}_j)}\langle S_i^zS_j^z\rangle,\\
 S^{\perp}_d(\bm{k})&=\frac{1}{N}\sum_{i,j}e^{-\iu \bm{k}\cdot(\bm{r}_i-\bm{r}_j)}\langle S_i^xS_j^x+S_i^yS_j^y\rangle,\\
 S_q^{\perp}(\bm{k})&=\frac{1}{N}\sum_{i,j}e^{-\iu \bm{k}\cdot(\bm{r}_i-\bm{r}_j)}\langle Q^{x^2-y^2}_iQ^{x^2-y^2}_j+ Q_i^{xy}Q_j^{xy}\rangle,
\end{align}
\end{subequations}
where $Q_i^{x^2-y^2}\equiv S_i^xS_i^x-S_i^yS_i^y$,
$Q_i^{xy}\equiv S_i^xS_i^y+S_i^yS_i^x$.
The labels $i,j$ run over the center
$\frac{1}{3}L_x\times L_y$ sites on a $L_x\times L_y$ cylinder,
and for a torus they run over all the sites. The results for the torus can be found in Fig.~5{\bf b} in the main text, which agree well with the perturbative calculations. Additional results on the cylinders are shown in Fig.~\ref{order1}, which also agree well with those in the torus. Due to the limited system sizes available, we did not attempt to
extrapolate the order parameters to the thermodynamic limit.

\end{document}